%% file: main.tex
\documentclass[journal]{IEEEtran}
\ifCLASSINFOpdf
\else
\fi
\usepackage{multirow} 

\usepackage{graphicx} 
\usepackage{amsmath} 
\usepackage{booktabs}
\usepackage{amssymb}

\begin{document}
%
\title{A Lightweight Model for Perceptual Image Compression via Implicit Priors}
%
%
%

\author{Hao~Wei,
        Yanhui~Zhou,
        Yiwen~Jia,
        Chenyang Ge,
        Saeed~Anwar,
        and~Ajmal~Mian

\thanks{Hao Wei, Yiwen Jia and Chenyang Ge are with National Key Laboratory of Human-Machine Hybrid Augmented Intelligence, Institute of Artificial Intelligence and Robotics, Xi'an Jiaotong University, Xi'an 710049, China (e-mail: haowei@stu.xjtu.edu.cn; jiayiwen@stu.xjtu.edu.cn; cyge@mail.xjtu.edu.cn).}

\thanks{Yanhui Zhou is with the School of Information and Telecommunication, Xi'an Jiaotong University, Xi'an 710049, China (e-mail: zhouyh@mail.xjtu.edu.cn).}


\thanks{Ajmal Mian and Saeed Anwar are with the Department of Computer Science and Software Engineering, The University of Western Australia, Perth, Crawley, WA 6009, Australia (e-mail:ajmal.mian@uwa.edu.au, saeed.anwar@uwa.edu.au). Ajmal Mian is the recipient of an Australian Research Council Future Fellowship Award (project number FT210100268) funded by the Australian Government.}
}

\maketitle

\input{01_abstract}
\input{02_introduction}
\input{03_related-work}

\input{04_methodology}
\input{05_experiments}
\input{06_conclusion}

\bibliographystyle{IEEEtran}
\bibliography{main}

\input{09_biography}

%
\IEEEpeerreviewmaketitle

\ifCLASSOPTIONcaptionsoff
  \newpage
\fi

\end{document}

%% file: 01_abstract.tex
\begin{abstract}

Perceptual image compression has shown strong potential for producing visually appealing results at low bitrates, surpassing classical standards and pixel-wise distortion-oriented neural methods.
However, existing methods typically improve compression performance by incorporating explicit semantic priors, such as segmentation maps and textual features, into the encoder or decoder, which increases model complexity by adding parameters and floating-point operations. This limits the model's practicality, as image compression often occurs on resource-limited mobile devices.
To alleviate this problem, we propose a lightweight perceptual \textbf{I}mage \textbf{C}ompression method using \textbf{I}mplicit \textbf{S}emantic \textbf{P}riors (\textbf{ICISP}). 
We first develop an enhanced visual state space block that exploits local and global spatial dependencies to reduce redundancy. 
Since different frequency information contributes unequally to compression, we develop a frequency decomposition modulation block to adaptively preserve or reduce the low-frequency and high-frequency information.
We establish the above blocks as the main modules of the encoder-decoder, and to further improve the perceptual quality of the reconstructed images, 
we develop a semantic-informed discriminator that uses implicit semantic priors from a pretrained DINOv2 encoder. 
Experiments on popular benchmarks show that our method achieves competitive compression performance and has significantly fewer network parameters and floating point operations than the existing state-of-the-art.
We will release the code and trained models.
\end{abstract}

\begin{IEEEkeywords}
Perceptual image compression, visual state space model, frequency decomposition, implicit semantic priors.
\end{IEEEkeywords}

%% file: 02_introduction.tex
\section{Introduction}
\label{introduction}
\IEEEPARstart{T}{he} volume of image data is rapidly increasing, with images often captured and stored on low-power devices like smartphones, tablets, cameras, and Internet of Things devices. Efficient on-device compression and decompression algorithms are crucial for storage, transmission, and visualization.

\begin{figure}[htbp]
\centering
\includegraphics[width=0.48\textwidth]{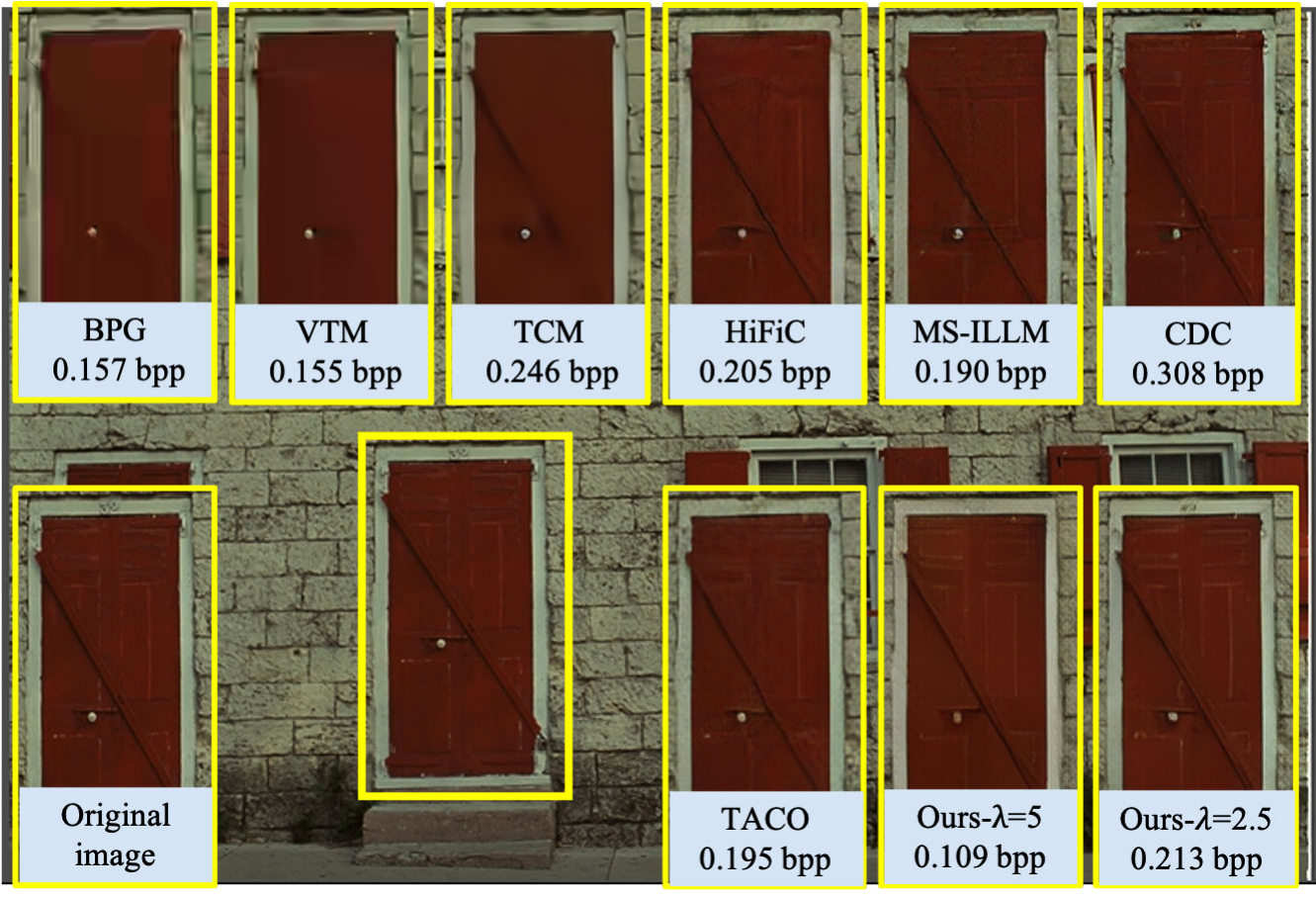} 
\caption{Visual comparison with state-of-the-art methods on the Kodak dataset shows that the proposed method achieves better reconstruction at lower bitrates with a well-preserved detailed door structure. Zoom in for the best view.}
\label{intro_visual}
\end{figure}

Classical standards such as JPEG~\cite{JPEG}, BPG~\cite{BPG}, and VVC~\cite{VVC} are widely used, but their block-wise processing often leads to artifacts, such as blockiness and blurriness, particularly at low bitrates.

Based on the variational auto-encoder (VAE) framework, many neural image compression methods using convolutional neural networks (CNN)~\cite{CNN_based, hyperprior} have been proposed. However, the local convolution operation limits their ability to capture long-range spatial dependencies, resulting in suboptimal compression performance. 
To alleviate this problem, several transformer-based image compression methods have been introduced. Transformers can effectively handle long contexts and outperform CNN-based approaches~\cite{TTC_ICML2021, TCM_CVPR2023, FAT_ICLR2024}. However, the quadratic computational complexity of self-attention in transformers limits their efficiency, and these methods often produce blurry results at low bitrates.

Motivated by generative adversarial networks (GAN)~\cite{GAN_ACMC2020}, some works design the decoder as a generator and introduce discriminators to help the compression network produce visually sharp results~\cite{GAN-ELIC_ICCV2019, HiFiC_NeurIPS2020, PO-ELIC_CVPR2022}. However, without appropriate priors, vanilla GAN-based methods often require huge models, especially at the decoder, which limits their practical use on memory- and power-constrained portable devices. For example, the decoder of the HiFiC method~\cite{HiFiC_NeurIPS2020} has 156.8 million parameters.

Several techniques have since improved perceptual compression performance by incorporating explicit semantic priors, such as semantic segmentation maps~\cite{DSSLIC_ICASSP2019} and textual descriptions~\cite{TGIC_AAAI2023, TACO_ICML2024}. However, adding semantic priors to the encoding or decoding process often results in large models with high computational complexity. For instance, the TACO method~\cite{TACO_ICML2024}, which uses text guidance, has 72.15 million parameters in its encoder and 7.34 million in its decoder, with a total of 328.9G FLOPs. Therefore, \textit{in this paper, we explore a lightweight perceptual compression method that incorporates semantic priors without increasing the complexity of the encoder or decoder.}

Inspired by the success of state space models that effectively capture the global context~\cite{Mamba_2023, VMamba_2024}, we first develop an enhanced visual state space block (EVSSB) to fully capture the local and the global spatial dependencies together. Since the contribution of information in images at different frequencies is not equal for compression, we further develop a frequency decomposition modulation block (FDMB) that adaptively selects the low/high-frequency information to be preserved or reduced. We integrate the proposed EVSSB and FDMB into the encoder and decoder of the proposed compression network, which can generate a compact feature representation, thus facilitating effective compression.

To further enhance the perceptual quality of the reconstructed images at low bitrates, we develop an effective semantic-informed discriminator that uses \emph{implicit} semantic priors from the pretrained DINOv2 encoder~\cite{DINOv2_TMLR2023}. 
Unlike existing prior-guided image compression methods, which embed \emph{explicit} semantic priors in the encoder/decoder~\cite{DSSLIC_ICASSP2019, TACO_ICML2024} leading to increased model size and computational complexity, we embed {\em implicit} model priors in the discriminator to cleverly avoid increasing the parameters and computational complexity of our encoder and decoder. Fig.~\ref{intro_visual} shows that our method is able to produce more realistic results at lower bitrates than other comparison methods. To summarize our contributions, we propose:
\begin{itemize}
    \item A lightweight model for image compression based on {\em implicit} semantic priors without adding extra parameters to the encoder or the decoder.
    \item An enhanced visual state space block (EVSSB) to capture both local and global spatial dependencies comprehensively, while a frequency decomposition modulation block (FDMB) that adaptively selects which low- and high-frequency information to retain or discard during compression.
    \item A semantic-informed discriminator using implicit semantic priors from the DINOv2 encoder to assist the compression network in semantically rich texture generation at low bitrates.     
\end{itemize}

The rest of this paper is organized as follows. The related works are summarized in Section \ref{related_works}. Section \ref{methodology} describes the details of the proposed method. The experimental results and analysis are presented in Section \ref{experiments}. Finally, we conclude our work in Section \ref{conclusion}.

%% file: 03_related-work.tex
\section{Related Works}
\label{related_works}

\subsection{Distortion-oriented Compression Methods}
Deep learning has been extensively used for lossy image compression. Ball$\acute{e}$ et al. proposed the first end-to-end CNN-based method and then improved it using VAE architecture and hyperprior model~\cite{CNN_based, hyperprior}. Minnen et al. introduced context modeling for more accurate entropy coding and utilized pixel-cnn~\cite{pixel-cnn_NeurIPS2016} for context prediction~\cite{JAHP_NeurIPS2018}. 
Subsequently, Generalized divisive normalization~\cite{GDN_ICLR2016} is able to enhance non-linearity due to its ability to model the local joint statistics of natural images. 
Furthermore, various innovative methods aim to improve the interaction between high-frequency and low-frequency features, for example, using octave convolution~\cite{octave_TMM_2021}, dynamic frequency filter~\cite{FDNet_TCSVT2024} or wavelet transform~\cite{WeConvene_ECCV2024}. 
Moreover, the attention mechanism is also widely used in learned image compression~\cite{SAttn_CVPR2020, WACNN_CVPR2022}. 
Recently, the Transformer has been successfully applied to many low-level vision tasks~\cite{FCN_IET2024} due to its non-local modeling capability, motivating many researchers to explore Transformer-based compression methods. Some works directly use the Swin Transformer~\cite{SwinT_ICCV2021} for image compression~\cite{TTC_ICML2021, WACNN_CVPR2022}. In~\cite{TCM_CVPR2023}, a parallel Transformer-CNN mixture block is proposed, combining the local modeling ability of CNNs with the non-local modeling ability of transformers. 
Li et al.~\cite{FAT_ICLR2024} proposed a frequency-aware transformer that computes self-attention based on different window sizes. 
However, these methods struggle with efficient attention computation due to the quadratic computational complexity of transformers. To strike a better balance between compression performance and efficiency, a visual state space model-based compression method is proposed in~\cite{MambaVC_2024}, leveraging the long-range dependency capture and efficiency of state space models. However, only using the 2D selective scanning process breaks dependencies between spatially local tokens.
\subsection{Perceptual Learned Compression Methods}

Perception-oriented image compression methods have recently gained popularity for producing realistic reconstructions at low bitrates. 
As a pioneering work, HiFiC is proposed, a GAN-based model demonstrating impressive compression performance at low bitrates~\cite{HiFiC_NeurIPS2020}. Muckley et al.~\cite{MS_ILLM_PMLR2023} improved a non-binary discriminator conditioned on quantized local image representation. Körber et al.~\cite{EGIC_ECCV2024} further adopted human-annotated semantic labels instead of codebook indices for discriminator. Jia et al. performed compression in the generative latent space~\cite{GLC_CVPR2024}, inspired by VQGAN~\cite{VQGAN_CVPR2021, VQIR_TCSVT2024}. Akbari et al.~\cite{DSSLIC_ICASSP2019} used the semantic segmentation maps as prior to guide the encoding and decoding process. In~\cite{TGIC_AAAI2023}, text semantic information was introduced as prior to assist GAN-based perceptual image compression.
Subsequent works have started to use the diffusion model to achieve more realistic reconstructions~\cite{PerCo_ICLR2023, CorrDiff_ICML2024}. In~\cite{CDC_NeurIPS2024}, Yang et al. used the diffusion model as a decoder conditioned on the compressed latent features. 
In~\cite{text_sketch_ICML2023}, the compressed sketch and text descriptions were used as conditions for pre-trained diffusion models. 
Ma et al.~\cite{CorrDiff_ICML2024} designed an end-to-end decoder to transmit privileged information, which can help correct the sampling process of the diffusion-based decoder. 
Kuang et al.~\cite{CGDM_ACMMM2024} developed a diffusion-based post-processing network for detailed representation learning. However, diffusion is a computationally expensive process that is unsuitable for resource-constrained devices.

From the above-mentioned methods, we note that using explicit semantic priors, for example, semantic segmentation maps~\cite{DSSLIC_ICASSP2019} and textual descriptions~\cite{TACO_ICML2024}, can assist perceptual image compression. However, using explicit semantic priors to guide the encoding and decoding significantly increases the computational burden and memory requirements. 
To alleviate this, we develop a lightweight perceptual image compression method that explores the rich implicit semantic priors from pretrained encoders to aid compression without increasing the computational complexity of the encoder or the decoder.

%% file: 04_methodology.tex
\section{Methodology}
\label{methodology}
\begin{figure}
\centering
\includegraphics[width=0.2\textwidth]{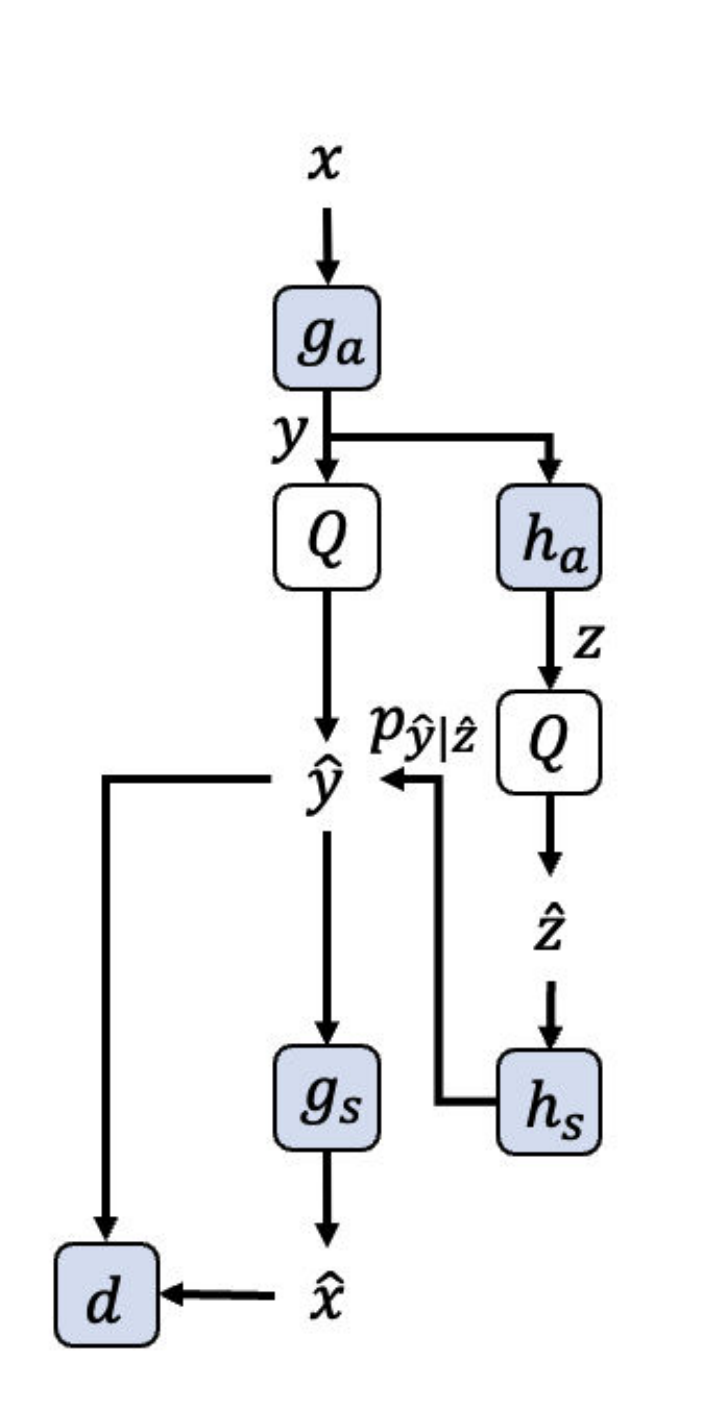} 
\includegraphics[width=0.2\textwidth]{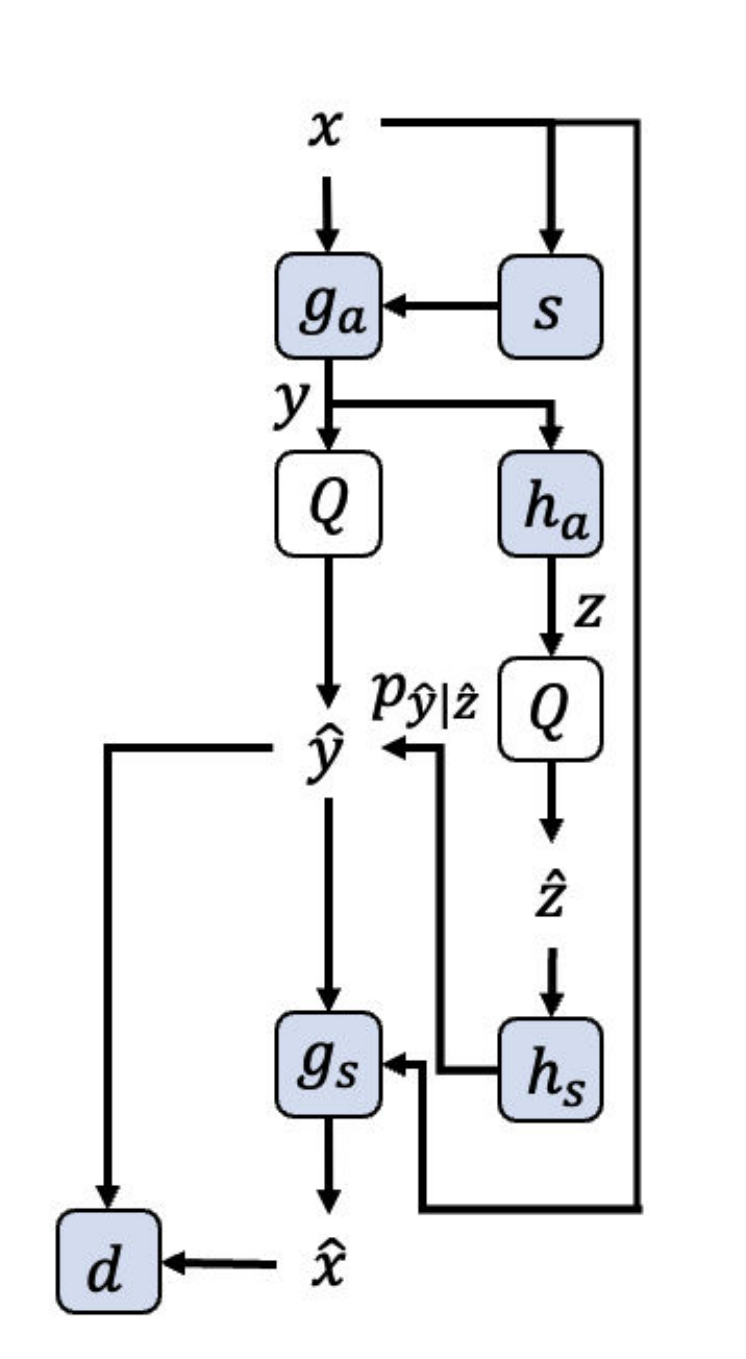}\\
\makebox[0.2\textwidth]{(a)}
\makebox[0.2\textwidth]{(b)}\\
\includegraphics[width=0.2\textwidth]{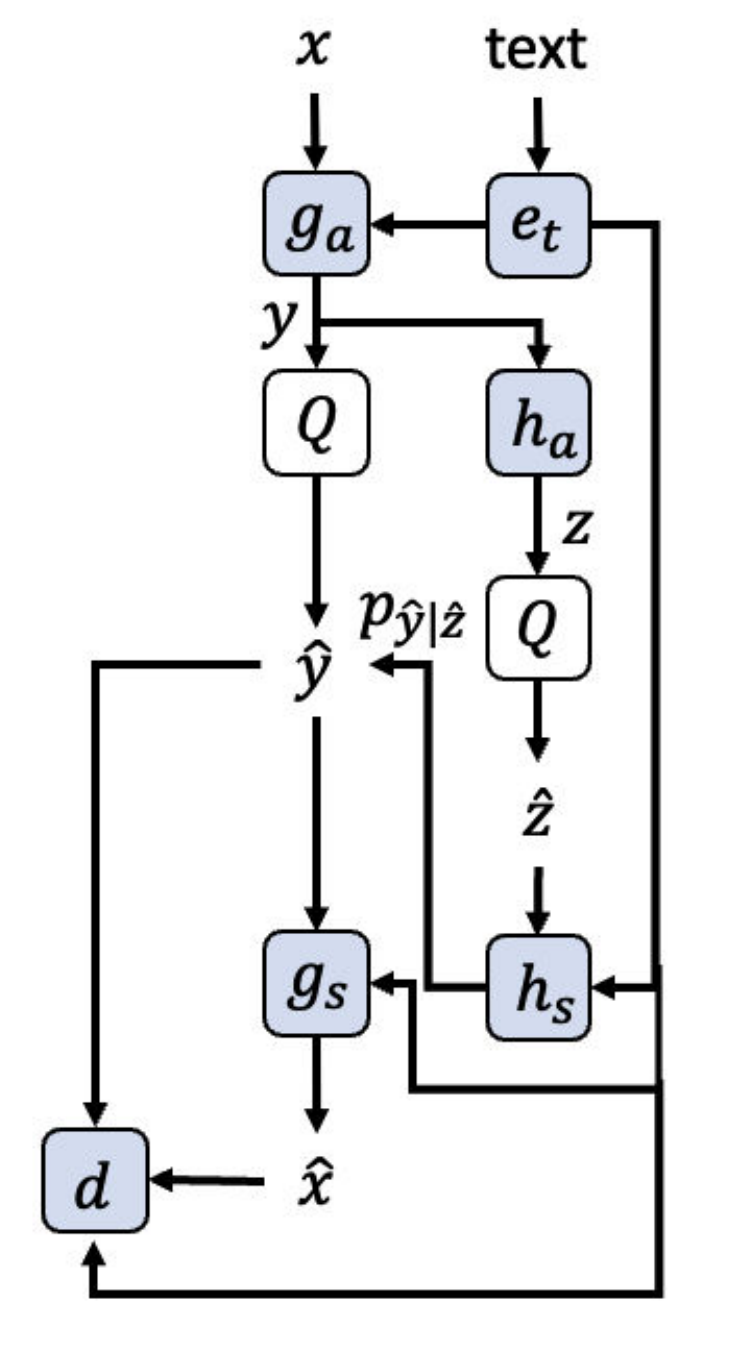} 
\includegraphics[width=0.2\textwidth]{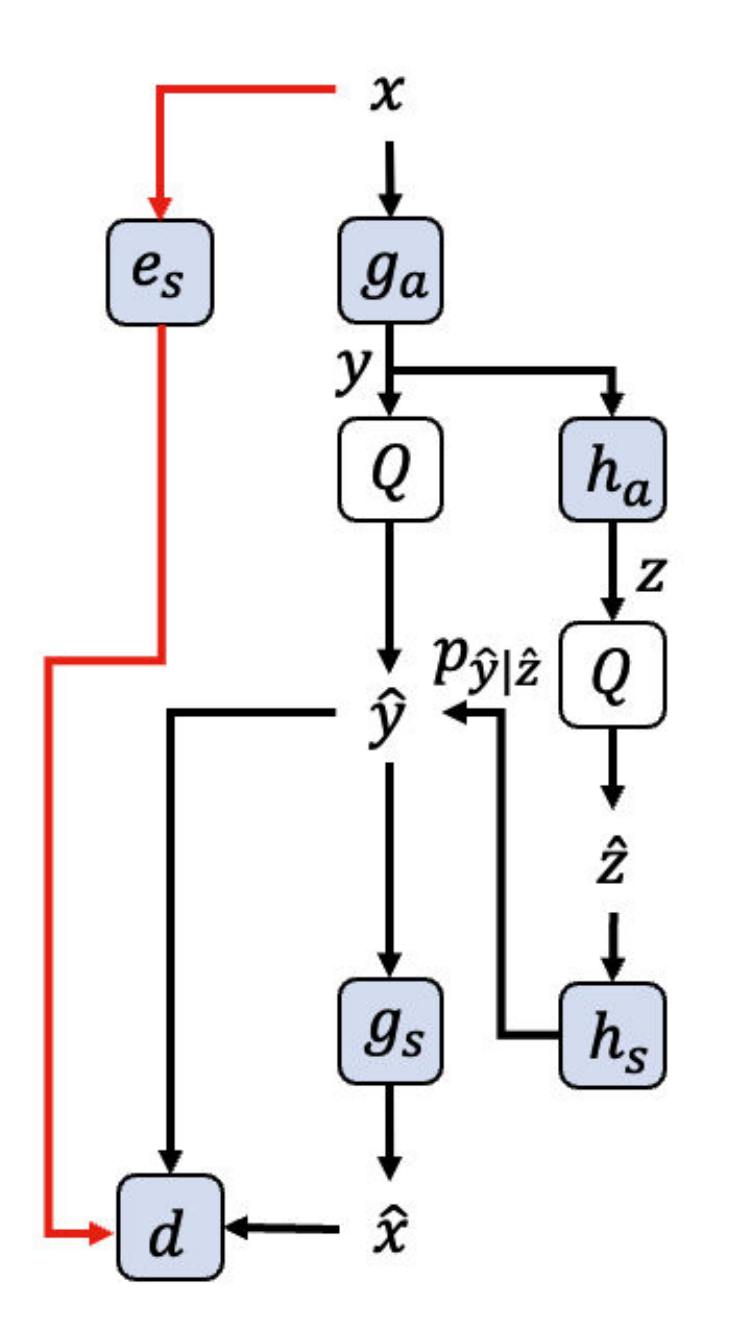} \\
\makebox[0.25\textwidth]{(c)}
\makebox[0.2\textwidth]{(d)}\\
\caption{Comparison of perceptual learned compression methods. (a) HiFiC~\cite{HiFiC_NeurIPS2020}. (b) DSLLIC using the explicit semantic segmentation map~\cite{DSSLIC_ICASSP2019}. (c) TGIC~\cite{TGIC_AAAI2023} using text as a guide. (d) Proposed method utilizing implicit semantic priors. Here, $s$, $e_t$, and $e_s$ represent the segmentation map generation, text encoder, and semantic priors generation, respectively.}
\label{framework_comparison}
\end{figure}

\subsection{Preliminary}
As shown in Fig.~\ref{framework_comparison}(a), the input image $\mathbf{x}$ is encoded into the latent $\mathbf{y}$ using the analysis transform $g_a$. Then $\mathbf{y}$ is quantized to $\mathbf{\hat{y}}$ and losslessly compressed into a bitstream using entropy coding, such as arithmetic coding. Finally, the synthesis transform $g_s$ reconstructs the image $\mathbf{\hat{x}}$. The whole process is formulated as:
\begin{equation}
\mathbf{y} = g_a(\mathbf{x}),\mathbf{\hat{y}} = Q(\mathbf{y}),\mathbf{\hat{x}} = g_s(\mathbf{\hat{y}}),
\end{equation}
where $Q(\cdot)$ is the quantization operation. 
To further capture the spatial dependencies in $\mathbf{\hat{y}}$, the hyper analysis transform $h_a$ is used to obtain the coded variables $\mathbf{z}$, which are further quantized to get their discrete version $\mathbf{\hat{z}}$. Following this, the hyper synthesis transform $h_s$ taking as input $\mathbf{\hat{z}}$ aims to estimate the parameters for the conditional probability distribution $p_{\mathbf{\hat{y}}|\mathbf{\hat{z}}}$.
To improve the perceptual quality of the reconstructed image $\mathbf{\hat{x}}$, the discriminator $d$ is introduced to constrain the perceptual consistency between $\mathbf{\hat{x}}$ and the input $\mathbf{x}$ using adversarial training.

Subsequent works \cite{DSSLIC_ICASSP2019, TGIC_AAAI2023} have further advanced perceptual image compression by leveraging informative semantic priors. For example, Akbari et al. (Fig.~\ref{framework_comparison}(b)) used the explicit semantic segmentation map as a guide for encoding and decoding, while Jiang et al. (Fig.~\ref{framework_comparison}(c)) employed text to assist compression. In contrast, we propose an efficient perceptual image compression framework that utilizes implicit semantic priors, integrating them into the discriminator. This approach helps the decoder to generate fine-grained semantic details while avoiding unnecessary computational complexity in the codec (see Fig.~\ref{framework_comparison}(d)).

\subsection{Proposed Method}
\begin{figure*}[htbp]
\centering
\includegraphics[width=1.0\linewidth]{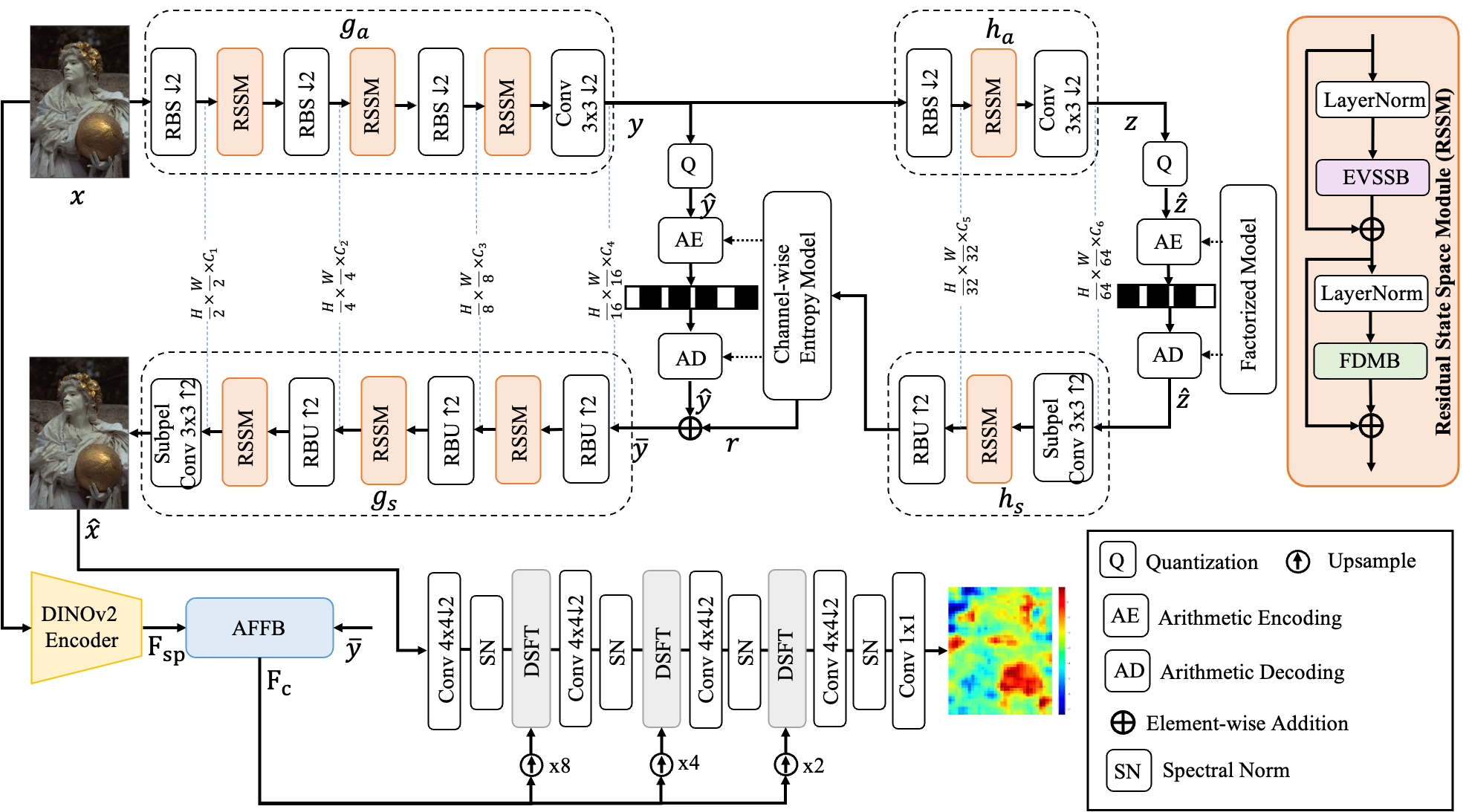}
\caption{Architecture overview. The proposed compression model uses residual block with stride (RBS), residual block upsampling (RBU), and residual state space module (RSSM) to build the nonlinear transforms ($g_a$, $h_a$, $h_s$ and $g_s$). The proposed semantic-informed discriminator exploits the semantic priors extracted from the DINOv2 encoder, which facilitates the semantic texture generation of the compression model at low bitrates. Details of the EVSSB and FDMB are in Fig.~\ref{arch_EVSSB} and~\ref{arch_FDMB}, whereas details of the AFFB and DSFT are in Fig.~\ref{AFFB_DSFT}. The RBS, RBU, and channel-wise entropy model are adapted from~\cite{SAttn_CVPR2020}.}
\label{framework}
\end{figure*}
Recall that our goal is to design an efficient perceptual image compression method that uses implicit semantic priors as a guide without adding computational complexity to the codec. To this end, we first develop an enhanced visual state space block to efficiently capture the local and non-local spatial dependencies for redundancy reduction. We further design a frequency decomposition modulation block to adaptively select the information to preserve or eliminate from a frequency perspective. We integrate the above blocks into the encoder and decoder of the compression network, and to further improve the perceptual quality of the reconstructed images, we propose a semantic-informed discriminator that uses the implicit semantic priors extracted from the pretrained encoder. 
Specifically, an adaptive feature fusion block is proposed to generate the effective condition based on implicit semantic priors and quantized latent features from the compressor. The conditioned features are then used to modulate the intermediate features of the discriminator via the proposed dynamic spatial feature transform, which improves the discriminative power and further facilitates the semantic texture generation of the compression network at low bitrates. 
Fig.~\ref{framework} provides an overview of the proposed method. Below, we describe the individual modules of the method in detail.

\subsection{Enhanced Visual State Space Block}
\begin{figure}
\centering
\includegraphics[width=0.48\textwidth]{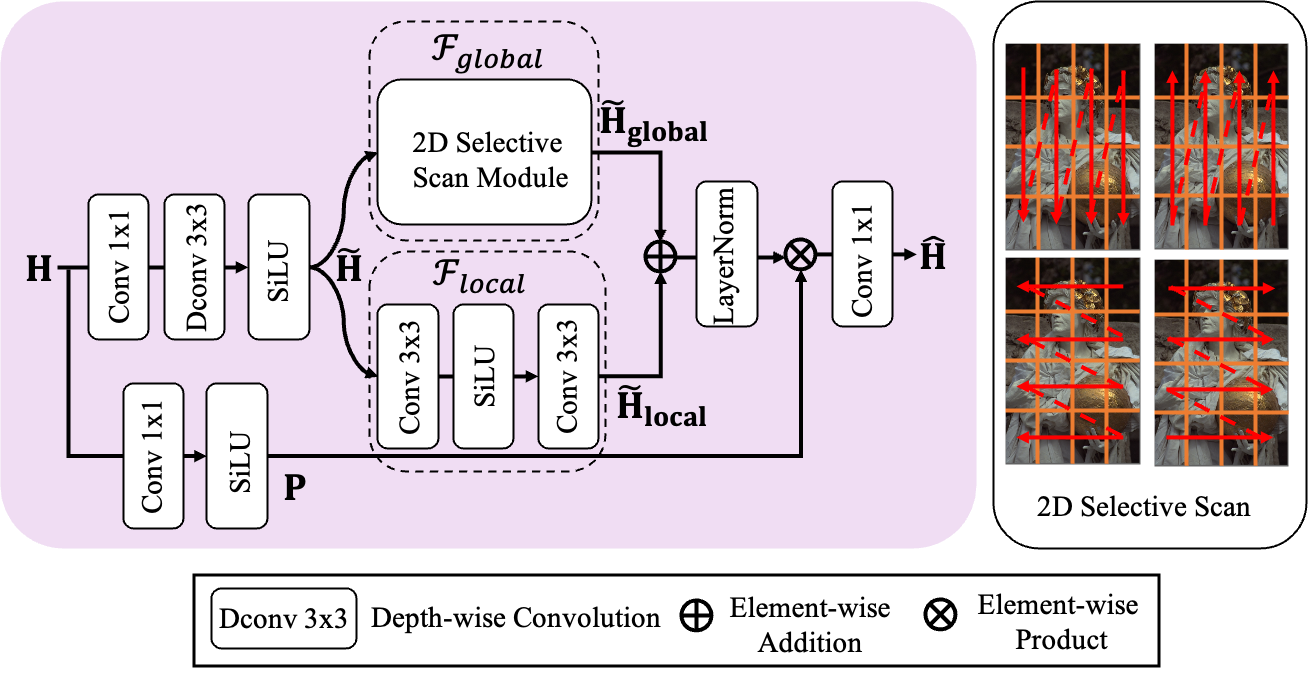} 
\caption{Network details of the proposed EVSSB.}
\label{arch_EVSSB}
\end{figure}

Capturing spatial dependencies is crucial for image compression. Recent works use CNNs, transformers, or a combination of both to explore spatial relationships for reducing redundant information~\cite{WACNN_CVPR2022, TTC_ICML2021, TCM_CVPR2023}. However, the local receptive fields of convolutions limit the model's feature representation, while the computational cost of dot-product attention is substantial, especially for high-resolution images.
In contrast, we propose an enhanced visual state space block (EVSSB), motivated by the VMamba~\cite{VMamba_2024}, to fully explore the spatial relationships of the features, both locally and globally. Unlike~\cite{MambaVC_2024}, we emphasize the local modeling branch in the proposed EVSSB to capture the spatial relationships between neighboring pixels.

Specifically, given the features $\mathbf{H}$, we first apply a 1$\times$1 convolution to aggregate the pixel-wise cross-channel context and the 3$\times$3 depth-wise convolution to extract the channel-wise spatial contexts $\mathbf{\tilde{H}}$. Then, we extract the global and local features independently in parallel by:
\begin{equation}
    \mathbf{\tilde{H}_{global}}=\mathcal{F}_{global}(\mathbf{\tilde{H}}), 
    \mathbf{\tilde{H}_{local}}=\mathcal{F}_{local}(\mathbf{\tilde{H}}),
\vspace{-1mm}
\end{equation}
where $\mathcal{F}_{global}(\cdot)$ and $\mathcal{F}_{local}(\cdot)$ is implemented by the 2D selective scan module~\cite{VMamba_2024} and two successive 3$\times$3 convolutions, respectively. Meanwhile, we introduce the gating branch to generate the gating parameters $\mathbf{P}$ as:
\begin{equation}
    \mathbf{P} = \mathcal{S}(\mathcal{F}_{conv}^{1\times1}(\mathbf{H})),
\vspace{-1mm}
\end{equation}
where $\mathcal{S}(\cdot)$ is SiLU activation and $\mathcal{F}_{conv}^{1\times1}(\cdot)$ denotes a 1$\times$1 convolution.
Finally, we obtain the enhanced features $\mathbf{\hat{H}}$ by:
\begin{equation}
    \mathbf{\hat{H}}=\mathcal{F}_{conv}^{1\times1}(\mathcal{F}_{ln}(\mathbf{\tilde{H}_{global}}+\mathbf{\tilde{H}_{local}})\otimes\mathbf{P}),
\end{equation}
where $\mathcal{F}_{ln}(\cdot)$ denotes the layer normalization, and $\otimes$ denotes the element-wise multiplication operation. The detailed network of the proposed EVSSB is shown in Fig.~\ref{arch_EVSSB}.

\subsection{Frequency Decomposition Modulation Block}
\begin{figure}
\centering
\includegraphics[width=0.48\textwidth]{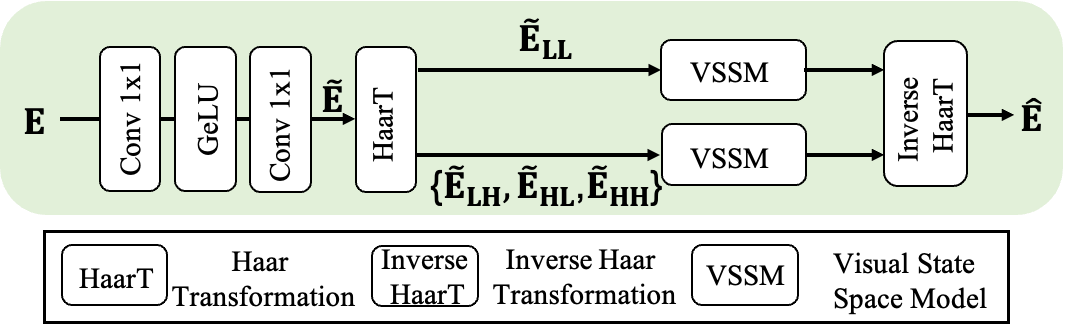} 
\caption{Network details of the proposed FDMB.}
\label{arch_FDMB}
\end{figure}
Note that EVSSB captures local and global dependencies in the spatial domain without accounting for the image's structures and details. It is well known that the contributions of these information types to image compression are not equal. For instance, compression models produce smooth results at low bitrates by relying on low-frequency information to reconstruct overall structures, while recovering fine details at high bitrates, which require more high-frequency information.

Inspired by this observation, we propose a frequency decomposition modulation block (FDMB), which adaptively selects the information to preserve or eliminate. As shown in Fig.~\ref{arch_FDMB},
given the features $\mathbf{E}$, we first adopt a regular feed-forward network to process each pixel location independently as:
\begin{equation}
    \mathbf{\tilde{E}}=\mathcal{F}_{conv}^{1\times1}(\mathcal{G}(\mathcal{F}_{conv}^{1\times1}(\mathbf{E}))),
\vspace{-1mm}
\end{equation}
where $\mathcal{G}(\cdot)$ denotes the GeLU activation. 
We then apply the Haar transform to decompose the intermediate features $\mathbf{\tilde{E}}$ into the low-frequency $\mathbf{\tilde{E}_{LL}}$ and high-frequency $\left\{ \mathbf{\tilde{E}_{LH},\mathbf{\tilde{E}_{HL}}, \mathbf{\tilde{E}_{HH}}}\right\}$ components, which are further modulated by two separate visual state space models (VSSM). Note that the VSSM is implemented by disabling the $\mathcal{F}_{local}(\cdot)$ of the EVSSB. Finally, we use the inverse Haar transform to reconstruct the features $\mathbf{\hat{E}}$. The entire process is formulated as:
\begin{equation}
\begin{split}
    \mathbf{\tilde{E}_{LL}}, \left\{ \mathbf{\tilde{E}_{LH},\mathbf{\tilde{E}_{HL}}, \mathbf{\tilde{E}_{HH}}}\right\}=\mathcal{H}(\mathbf{\tilde{E}}),\\
    \mathbf{\hat{E}}=\mathcal{H}^{-1}(\mathcal{F}_{v}(\mathbf{\tilde{E}_{LL}}),\mathcal{F}_{v}(\mathcal{C}(\mathbf{\tilde{E}_{LH}},\mathbf{\tilde{E}_{HL}},\mathbf{\tilde{E}_{HH}}))),
\end{split}
\end{equation}
where $\mathcal{F}_{v}(\cdot)$ denotes the VSSM, $\mathcal{C}(\cdot)$ is the channel-wise concatenation operation, and $\mathcal{H}(\cdot)$ and $\mathcal{H}^{-1}(\cdot)$ denote the Haar transform and its inverse version, respectively.

\subsection{Semantic-informed Discriminator}
To enable the compression network to generate fine-grained, semantic-related textures at low bitrates, we propose a semantic-informed discriminator that fully leverages implicit semantic priors from a pretrained vision model. This approach addresses three key aspects: 1) How to extract semantic features from images? 2) How to effectively fuse the semantic information and the latent $\mathbf{\bar{y}}$? 3) How to incorporate the fused features as conditions into the discriminator?
\begin{figure}[htbp]
\centering
\includegraphics[width=0.48\textwidth]{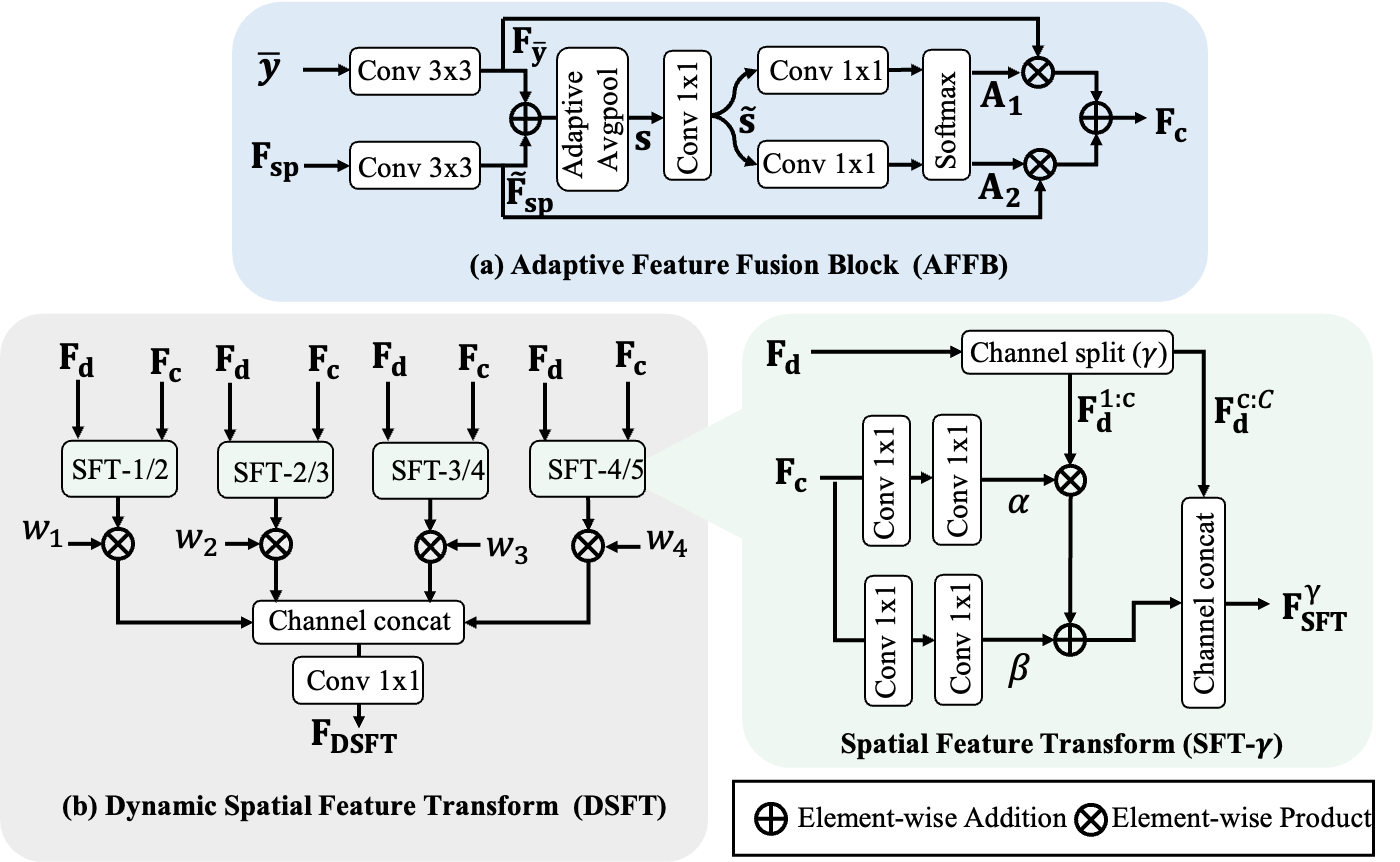} 
\caption{Network details of the proposed AFFB and DSFT.}
\label{AFFB_DSFT}
\end{figure}

\subsubsection{Semantic extraction} Pretrained vision models, leveraging large-scale datasets~\cite{ImageNet22K}, have proven their potential across many vision tasks, including segmentation~\cite{Humanseg_CVPR2024} and generation~\cite{LDM_CVPR2022}. Therefore, we resort to the DINOv2 model, which excels in semantic-related tasks~\cite{depth_anything_CVPR2024}. Specifically, we apply the DINOv2 encoder to the original image $\mathbf{x}$ for semantic-aware feature extraction as:
\begin{equation}
    \mathbf{F_{sp}}=\mathcal{N}_{sp}(\mathbf{x}),
\vspace{-1mm}
\end{equation}
where $\mathcal{N}_{sp}(\cdot)$ is the frozen pretrained DINOv2 encoder and $\mathbf{F_{sp}}$ are semantic features.

\subsubsection{Adaptive feature fusion} Motivated by~\cite{SKNet_CVPR2019}, we propose an adaptive feature fusion block (AFFB) to adaptively fuse the semantic features and the latent, as shown in Fig.~\ref{AFFB_DSFT}(a).
Specifically, given the semantic features $\mathbf{F_{sp}}$ and the latent $\mathbf{\bar{y}}$, we first apply a convolution layer to each independently as:  
\begin{equation}
    \mathbf{F_{\bar{y}}}=\mathcal{F}_{conv}^{3\times3}(\mathbf{\bar{y}}), ~~~~~
    \mathbf{\tilde{F}_{sp}}=\mathcal{F}_{conv}^{3\times3}(\mathbf{F_{sp}}),
\vspace{-1mm}
\end{equation}
where $\mathbf{F_{\bar{y}}}$ and $\mathbf{\tilde{F}_{sp}}$ are enhanced latent and semantic features, respectively, with the same dimension. We then combine them using element-wise summation, followed by global adaptive average pooling to generate channel-wise statistics $\mathbf{s}$. Next, a 1$\times$1 convolution compresses these statistics into compact features $\mathbf{\tilde{s}}$ with a squeezing factor of 2. Following this, two parallel convolutions expand $\mathbf{\tilde{s}}$, and a softmax function is applied to obtain attention vectors $\mathbf{A_1}$ and $\mathbf{A_2}$. Finally, the fused features $\mathbf{F_c}$ are generated through adaptive aggregation as:
\begin{equation}
    \mathbf{F_c} = \mathbf{A_1}\otimes\mathbf{F_{\bar{y}}}+\mathbf{A_2}\otimes\mathbf{\tilde{F}_{sp}}.
\end{equation}

\subsubsection{Dynamic spatial feature transform} To integrate the fused features into the discriminator, we use Spatial Feature Transform (SFT)~\cite{GFPGAN_CVPR2021}, which generates affine transformation parameters to modulate the discriminator's intermediate features. As shown in Fig.~\ref{AFFB_DSFT}, given the discriminator features $\mathbf{F_d}$ with the channel dimension of $C$, we first split $\mathbf{F_d}$ into two independent parts $(\mathbf{F}^{1:c}_\mathbf{d},\mathbf{F}^{c:C}_\mathbf{d})$ along the channel dimension using the ratio $\gamma$. We then apply several convolutions to condition $\mathbf{F_c}$ and obtain the affine transformation parameters $(\alpha,\beta)$. Finally, the feature modulation process is formulated as:
\begin{equation}
    \mathbf{F^\gamma_{SFT}}=\mathcal{C}(\alpha\otimes\mathbf{F}^{1:c}_\mathbf{d}+\beta, \mathbf{F}^{c:C}_\mathbf{d}), c=\gamma\times C,
\vspace{-1mm}
\end{equation}
where $\mathcal{C}(\cdot)$ denotes the channel-wise concatenation operation. To further improve the discriminative ability of our discriminator, we propose a dynamic spatial feature transform using the multiple values of $\gamma$:
\begin{equation}
    \mathbf{F_{DSFT}} = \mathcal{F}_{conv}^{1\times1}(\mathcal{C}(\left \{w_{i}\otimes\mathbf{F}^{\frac{i}{i+1}}_\mathbf{SFT} \right \}_{i=1}^{4})),
\end{equation}
where $\left\{w_i\right\}^{4}_{i=1}$ are learnable parameters (see Fig.~\ref{AFFB_DSFT}(b)).
\subsection{Two-stage Training}
We adopt the two-stage training strategy for the proposed method. In the first stage, the discriminator is disabled, and the compression model is trained for rate-distortion optimization. The loss function for this stage is defined as:
\begin{equation}
\label{stageI}
    \mathcal{L}_{stage I} = \mathcal{R}(\mathbf{\hat{y}})+\mathcal{R}(\mathbf{\hat{z}})+k\cdot\mathcal{D}(\mathbf{x}, \mathbf{\hat{x}})
\end{equation}
where $\mathcal{R}(\mathbf{\hat{z}})$, $\mathcal{R}(\mathbf{\hat{y}})$ denote the bitrates of latent $\mathbf{\hat{z}}$ and $\mathbf{\hat{y}}$, respectively. $\mathcal{D}(\mathbf{x}, \mathbf{\hat{x}})$ represents the distortion loss computed using MSE. The parameter $k$ is set to 0.0067. In the second stage, we fine-tune the compression model by introducing the proposed semantic-informed discriminator to balance the rate-distortion-perception tradeoff. The supervision includes the rate loss $\mathcal{R}$, distortion loss $\mathcal{D}$, perceptual loss $\mathcal{L}_{per}$~\cite{LPIPS_CVPR_2018}, style loss $\mathcal{L}_{sty}$~\cite{styleloss_CVPR2016}, and adversarial loss $\mathcal{L}_{adv}$~\cite{PO-ELIC_CVPR2022}, as follows:
\begin{equation}
\label{stageII}
    \mathcal{L}_{stageII}=\lambda\cdot\mathcal{R}+k_1\cdot\mathcal{D}+k_2\cdot\mathcal{L}_{per}+k_3\cdot\mathcal{L}_{sty}+k_4\cdot\mathcal{L}_{adv},
\end{equation}
where the rate loss and distortion loss are equal to (\ref{stageI}); The perceptual loss $\mathcal{L}_{per}$ is defined as:
\begin{equation}
    \mathcal{L}_{per}=\|\psi(\mathbf{\hat{x}})-\psi(\mathbf{x})\|_{2},
\end{equation}
where $\psi(\cdot)$ denotes pretrained VGG network; The style loss $\mathcal{L}_{sty}$ is expressed as:
\begin{equation}
    \mathcal{L}_{sty}=\|G(\psi(\mathbf{\hat{x}}))-G(\psi(\mathbf{x}))\|_{1},
\end{equation}
where $G(\cdot)$ is the Gram matrix of the given features; The adversarial loss $\mathcal{L}_{adv}$ is defined as:
\begin{equation}
    \mathcal{L}_{adv}= -\mathbb{E}[\mathcal{F}_{d}(\mathbf{\hat{x}}, \mathbf{\bar{y}}, \mathbf{x})],
\end{equation}
where $\mathcal{F}_{d}(\cdot)$ is the proposed semantic-informed discriminator. The parameters $\{\lambda, k_1, k_2, k_3, k_4\}$ are given in Section \ref{implemental_details}. 

To constrain the proposed discriminator, we apply the hinge loss as:
\begin{equation}
    \mathcal{L}_{d}=-\mathbb{E}[\rho(-1+\mathcal{F}_{d}(\mathbf{x}, \mathbf{\bar{y}}, \mathbf{x})]-\mathbb{E}[\rho(-1-\mathcal{F}_{d}(\mathbf{\hat{x}}, \mathbf{\bar{y}}, \mathbf{x})],
\end{equation}
where $\rho(\cdot)$ is the ReLU function.


%% file: 05_experiments.tex
\section{Experiments}
\label{experiments}
\subsection{Settings and Datasets}
\subsubsection{Dataset} 
We train the proposed ICISP on the LSDIR dataset~\cite{LSDIR_CVPR2023}, which contains 84,991 high-quality images. After training, we evaluate our method on the widely used datasets, including the Kodak dataset~\cite{Kodak}, the CLIC2020 dataset~\cite{CLIC2020}, and the DIV2K dataset~\cite{DIV2K}.
The Kodak dataset consists of 24 natural images with a resolution of 768$\times$512, while the CLIC2020 dataset contains 428 images with resolutions up to 2000$\times$1000 pixels. The DIV2K dataset includes 100 images with 2K resolution.

\subsubsection{Evaluation Metrics} 
We measure bitrates in bits per pixel (bpp) and evaluate the reconstruction visual quality using perceptual metrics: LPIPS~\cite{LPIPS_CVPR_2018} and DISTS~\cite{DISTS_TPAMI_2020} for reference, and FID~\cite{FID} and KID~\cite{KID} for non-reference evaluation. Additionally, we use PSNR and MS-SSIM~\cite{MSSSIM} as reference metrics to assess distortion. Note that we do not include the FID/KID values comparison for the Kodak dataset, as its small size prevents the computation of FID/KID based on 299$\times$299 patches, as required by MS-ILLM~\cite{MS_ILLM_PMLR2023}.

\subsubsection{Implementation Details}
\label{implemental_details}
The proposed method is trained on a single NVIDIA GeForce RTX 4090 GPU. During training, we apply random cropping to the images, setting the spatial size of 256$\times$256, with a batch size of 8. 
In the first stage, the compression model is trained for 120 epochs using the Adam optimizer with Eq.(\ref{stageI}) and an initial learning rate of $10^{-4}$. The learning rate is reduced by half at epochs 110 and 115. 
In the second stage, we fine-tune the compression model using Eq.(\ref{stageII}) and train the semantic-informed discriminator from scratch for 50 epochs. The Adam optimizer is used for both the compression network and the discriminator with learning rates of $10^{-4}$. 
For the compression network, the learning rate is halved at epochs 40 and 45, while for the discriminator, it is halved at epochs 30 and 40. 
In Eq.(\ref{stageII}), the hyperparameters $k_1$, $k_2$, $k_3$ and $k_4$ are empirically set to 0.0004, 5, 2000 and 0.8, respectively. The value of $\lambda$ is selected from $\{1, 1.5, 2.5, 5, 7.5\}$. 
In addition, for the compression model, we set the intermediate features $\{C_1, C_2, C_3, C_4, C_5, C_6\}$ as $\{64, 64, 64, 320, 64, 192 \}$.
%
\begin{figure*}
\centering
\includegraphics[width=1\textwidth]{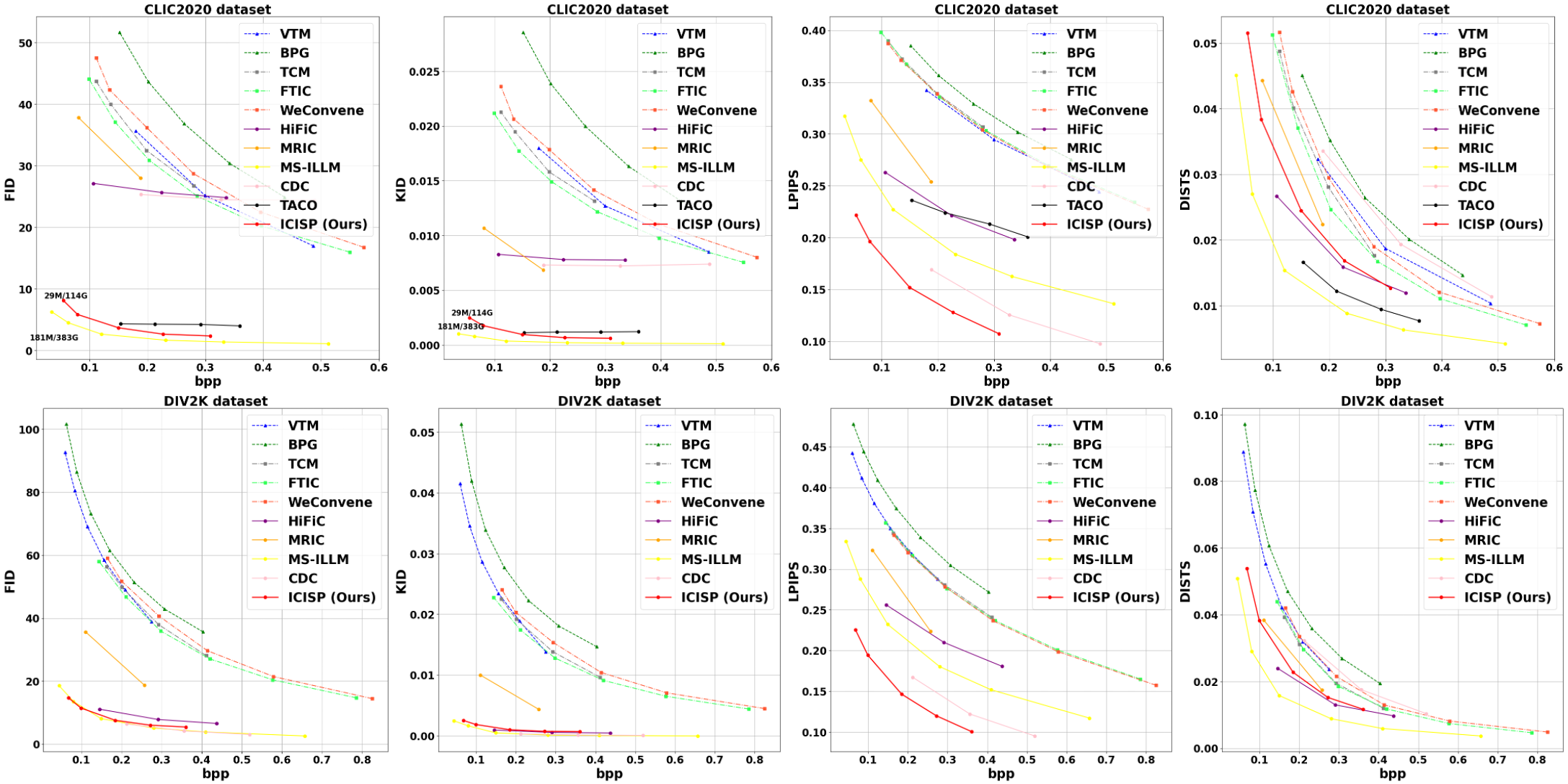}
\includegraphics[width=1\textwidth]{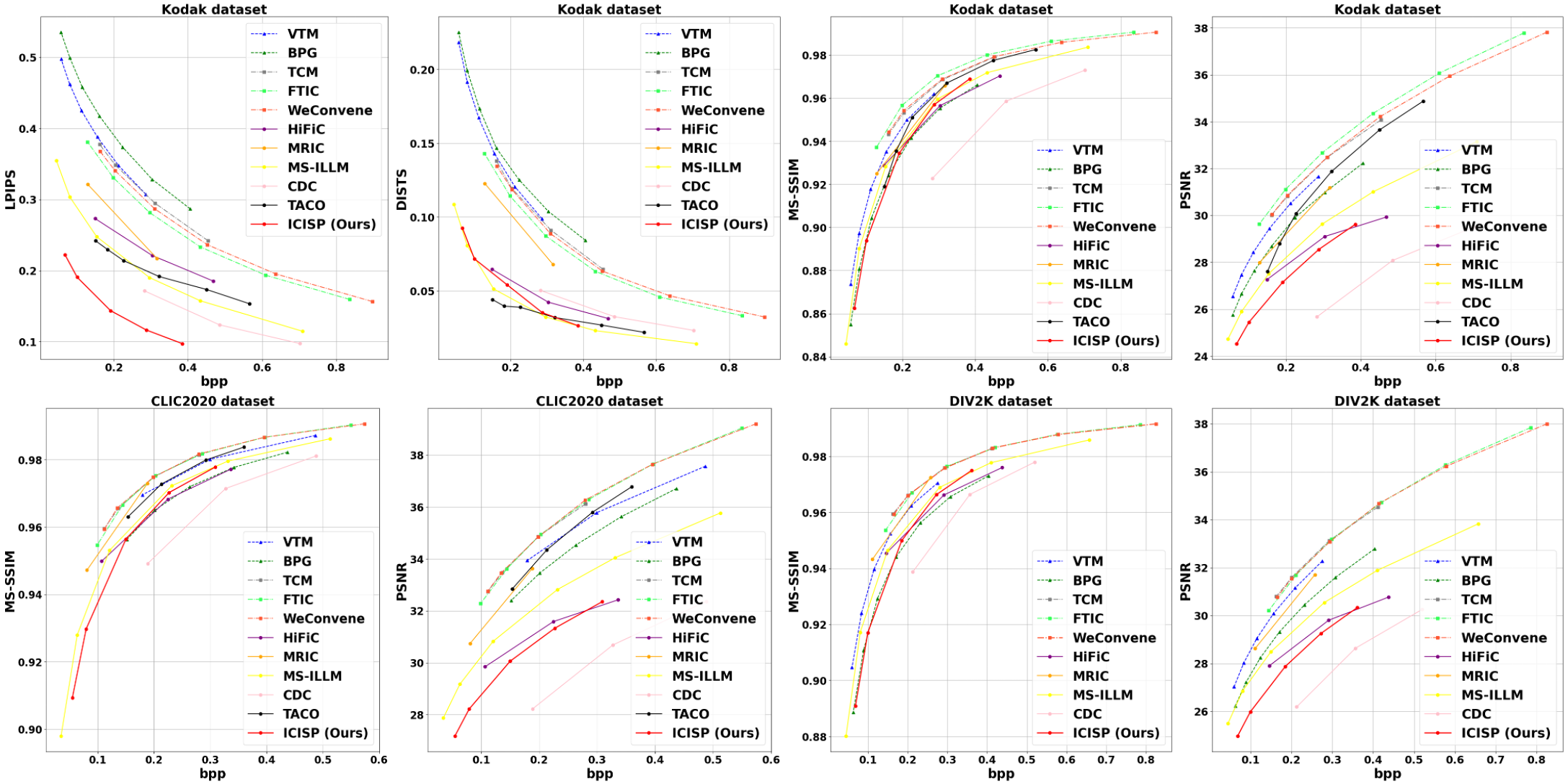}
\caption{Quantitative comparisons on the benchmark datasets. Lower FID/KID/LPIPS/DISTS and higher PSNR/MS-SSIM values are desirable as well as low bpp (bits per pixel). The parameters (M)/FLOPs (G) values are provided in the figures, with FLOPs calculated based on a 768×512 image patch. Zoom in for best view.}
\label{quan_com}   
\end{figure*}
\subsection{Comparisons with State-of-the-Art Methods}
We compare the proposed method with ten compression methods including traditional codecs (VTM~\cite{VVC} and BPG~\cite{BPG}), distortion-oriented codecs (TCM~\cite{TCM_CVPR2023}, FTIC~\cite{FAT_ICLR2024}, and WeConvene~\cite{WeConvene_ECCV2024}), and perceptual compression methods (HiFiC~\cite{HiFiC_NeurIPS2020}, MRIC~\cite{MRIC_CVPR2023}, MS-ILLM~\cite{MS_ILLM_PMLR2023}, CDC~\cite{CDC_NeurIPS2024}, and TACO~\cite{TACO_ICML2024}).
\subsubsection{Main results}
Fig. \ref{quan_com} shows quantitative comparisons on the benchmark datasets. As shown, the proposed ICISP outperforms traditional codecs and distortion-oriented methods in all perceptual metrics. When compared to other perceptual compression methods, our approach achieves the lowest LPIPS and comparable FID/KID values. While trailing MS-ILLM in DISTS, our method is much lighter, with significantly fewer parameters and FLOPs (see Table \ref{complexity_comparison}). In addition, it is observed that our method does not outperform in rate-distortion performance, as indicated by the PSNR/MS-SSIM metrics. This is because, during training with Eq. (\ref{stageII}), we prioritize the rate-perceptual tradeoff and do not explicitly optimize for distortion minimization, as we use a much smaller hyperparameter for the distortion loss compared to the perceptual loss.

\begin{figure*}
\centering
\includegraphics[width=0.24\textwidth]{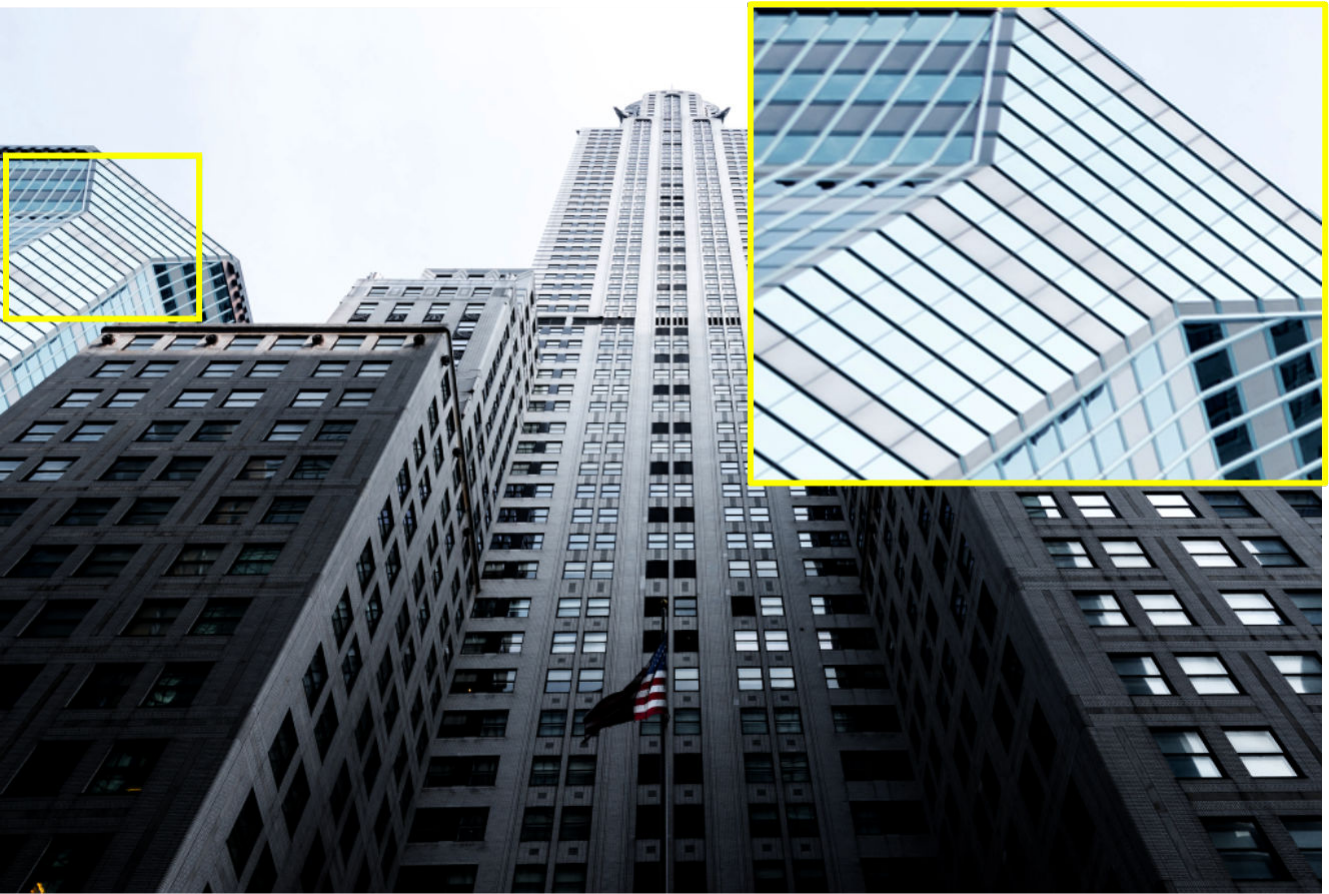} 
\includegraphics[width=0.24\textwidth]{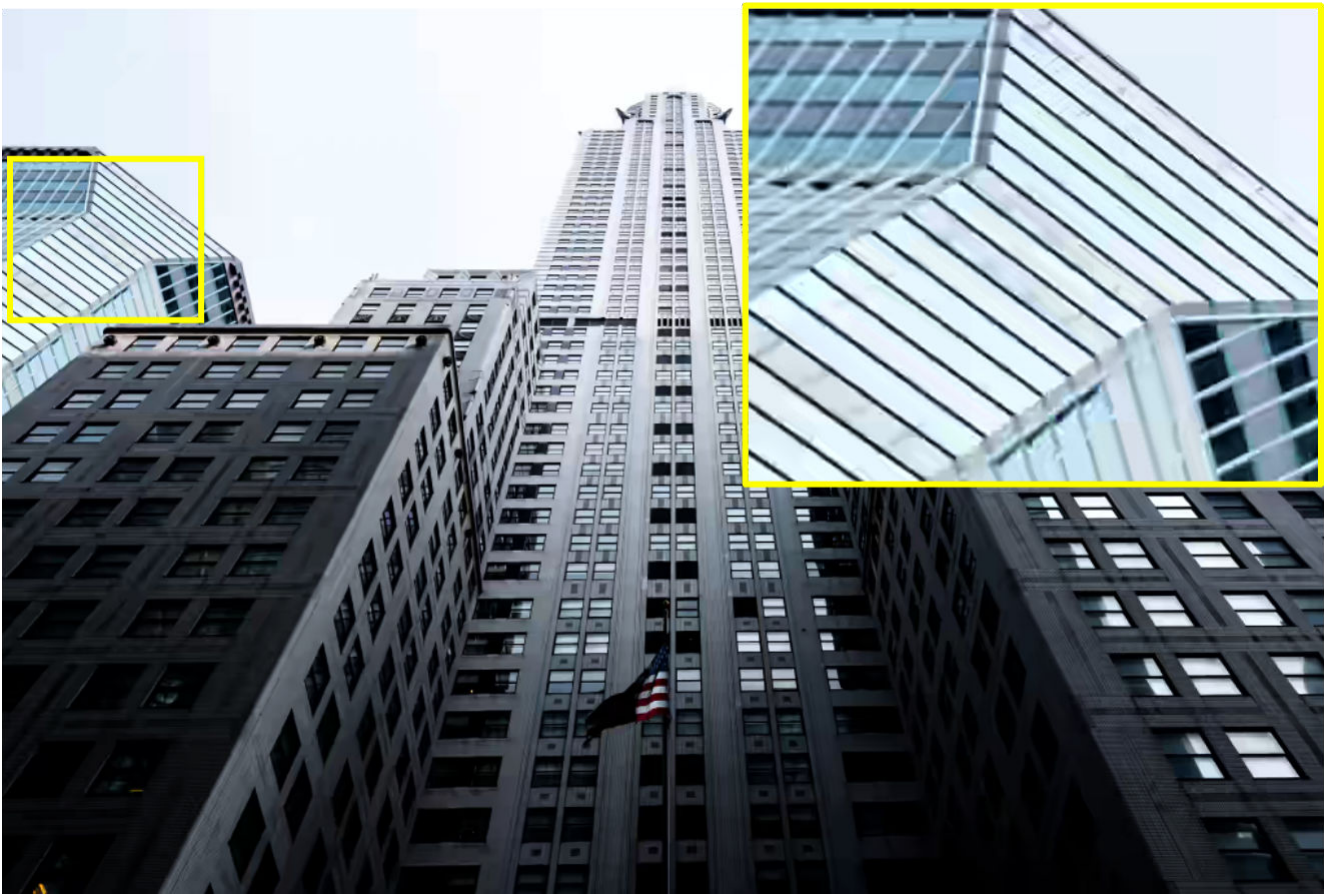} 
\includegraphics[width=0.24\textwidth]{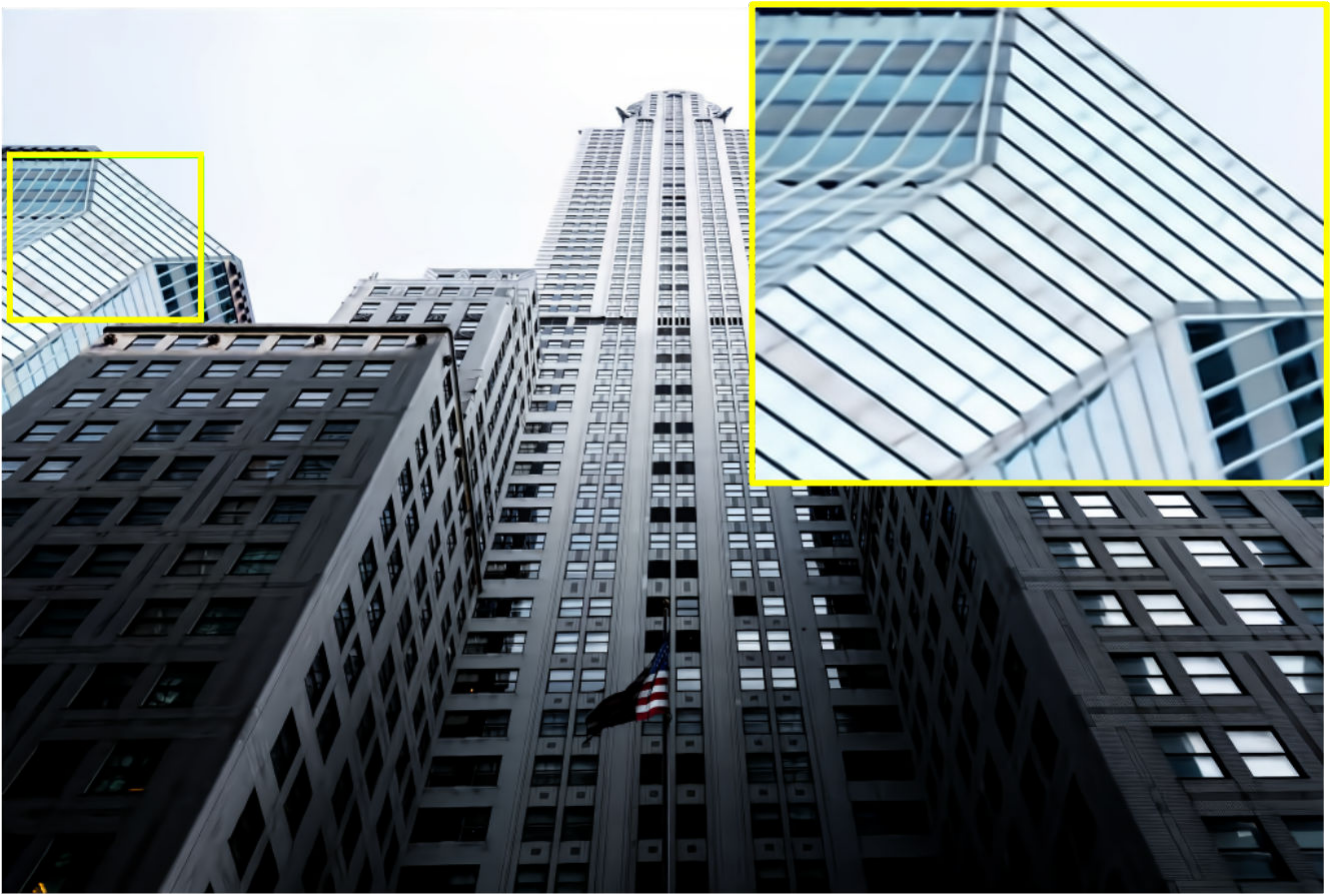} 
\includegraphics[width=0.24\textwidth]{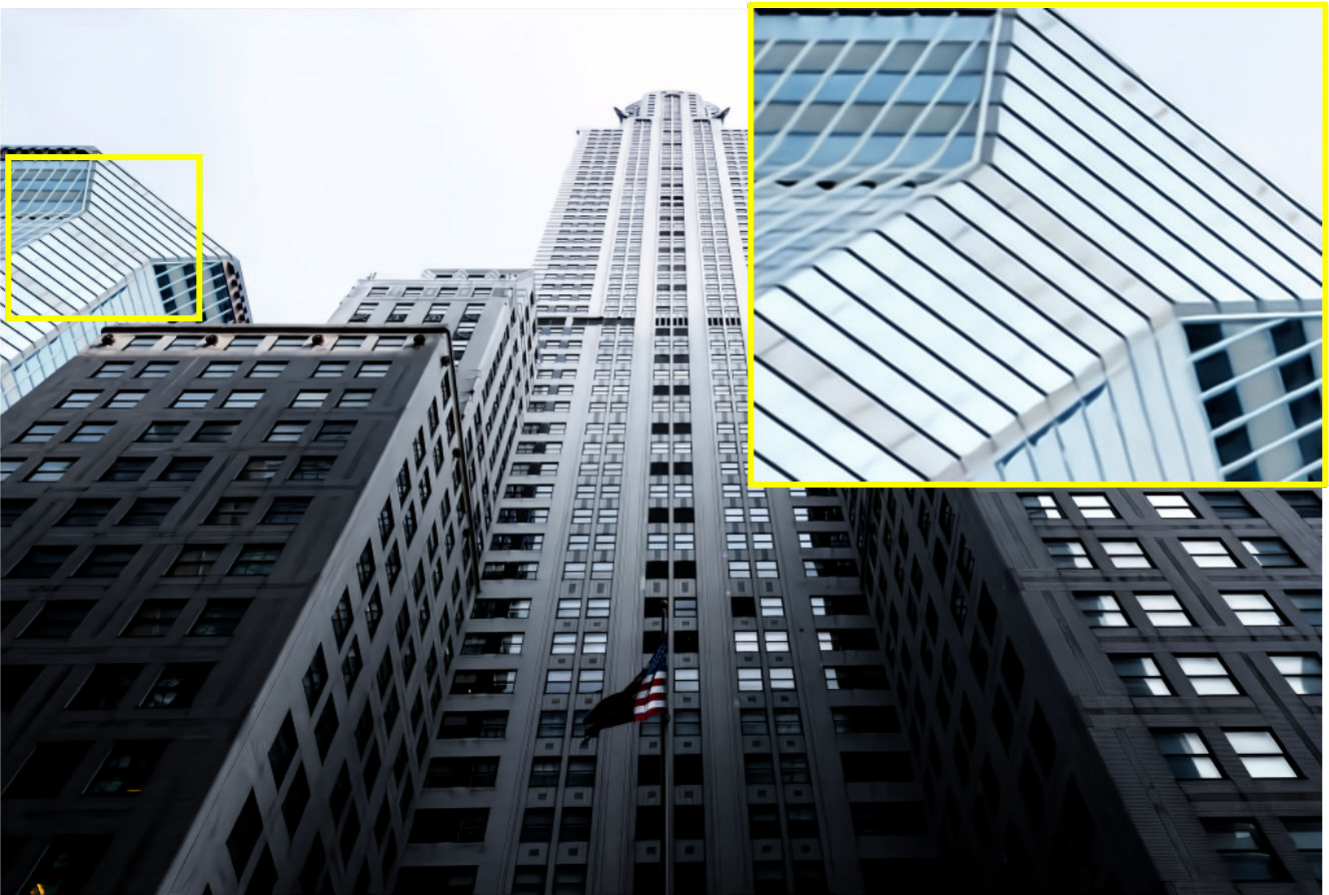} \\
\makebox[0.24\textwidth]{Original image (bpp)}
\makebox[0.24\textwidth]{BPG (0.134)~\cite{BPG}}
\makebox[0.24\textwidth]{TCM (0.140)~\cite{TCM_CVPR2023}} 
\makebox[0.24\textwidth]{FTIC (0.117)~\cite{FAT_ICLR2024}}\\
\includegraphics[width=0.24\textwidth]{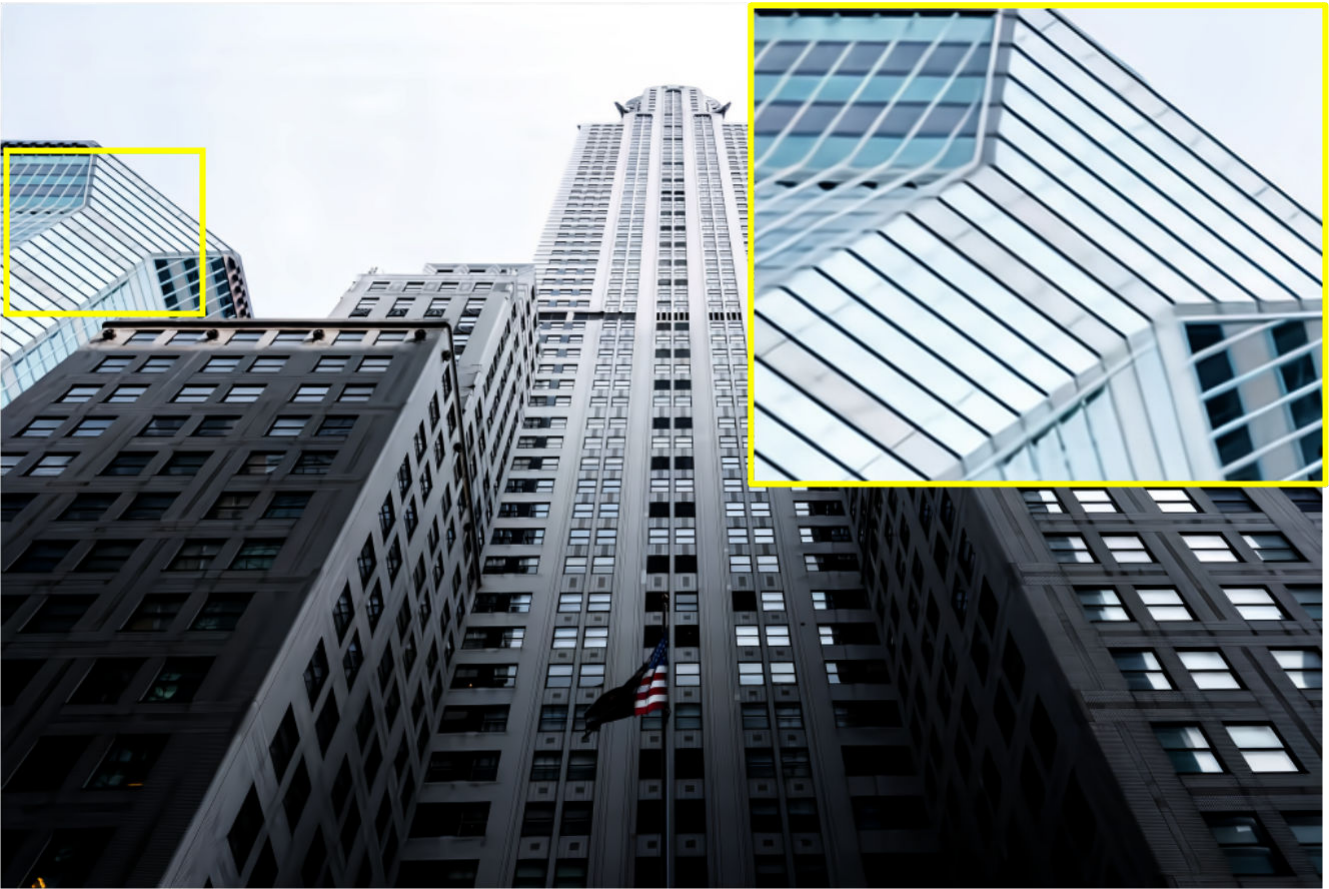}
\includegraphics[width=0.24\textwidth]{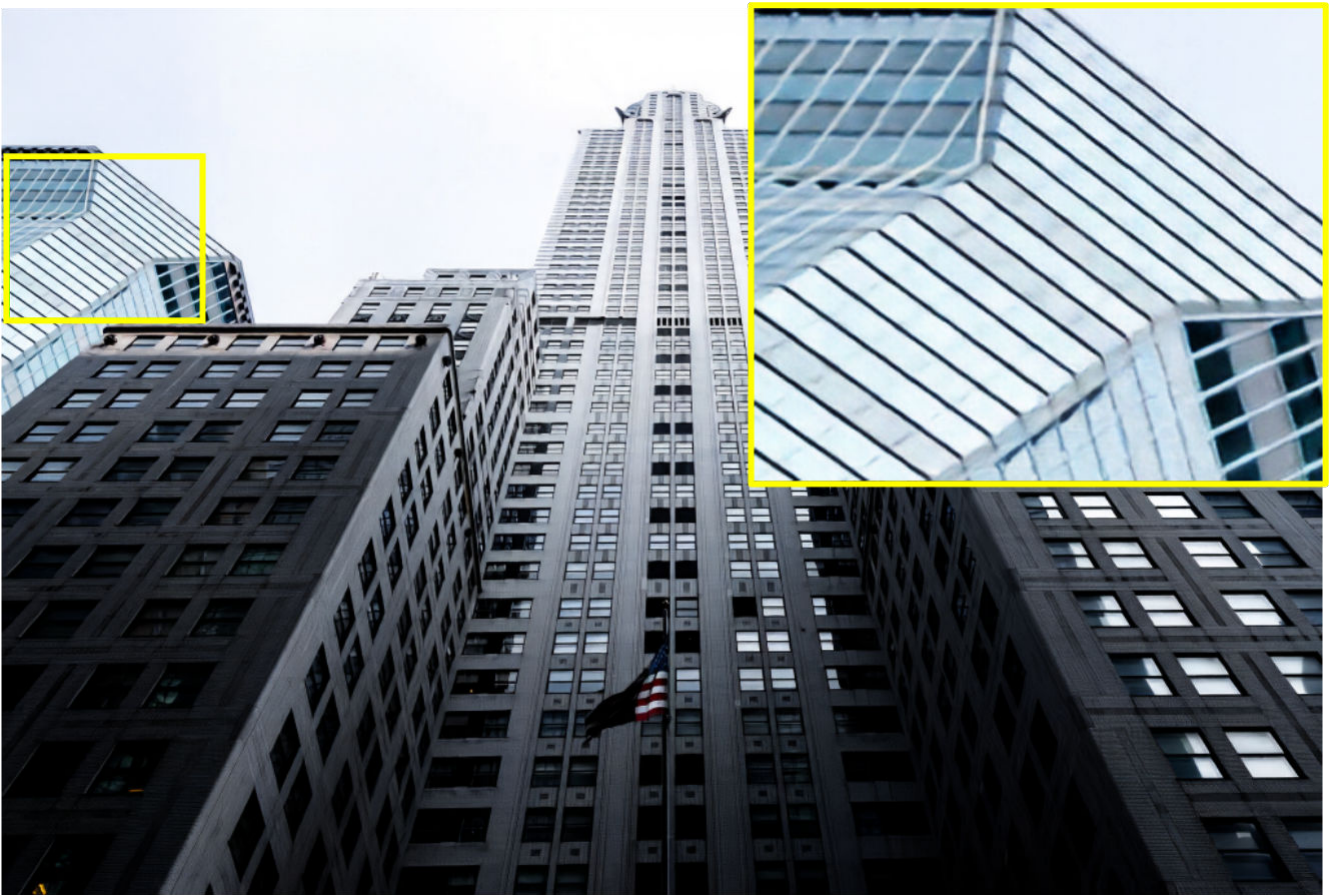} 
\includegraphics[width=0.24\textwidth]{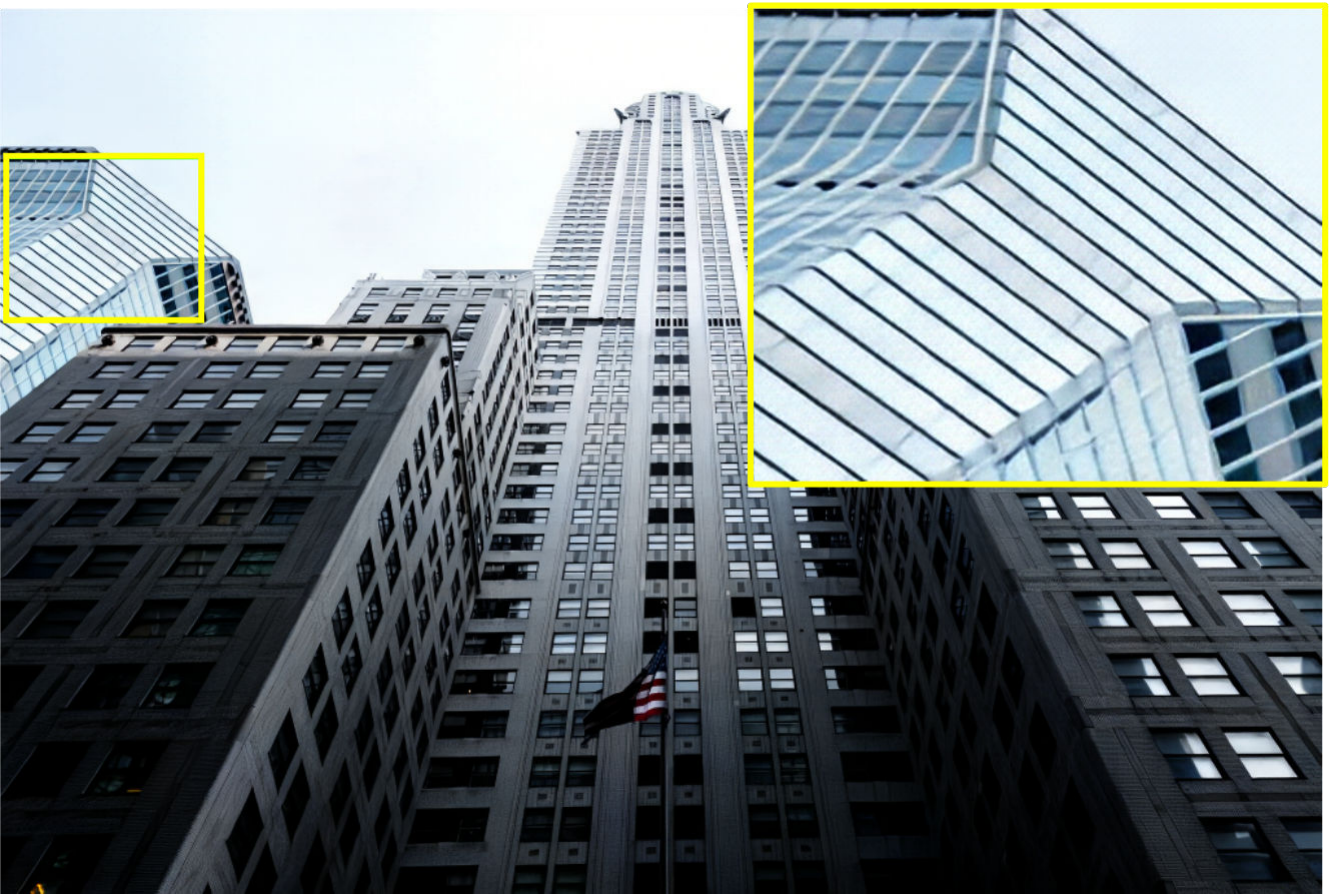} 
\includegraphics[width=0.24\textwidth]{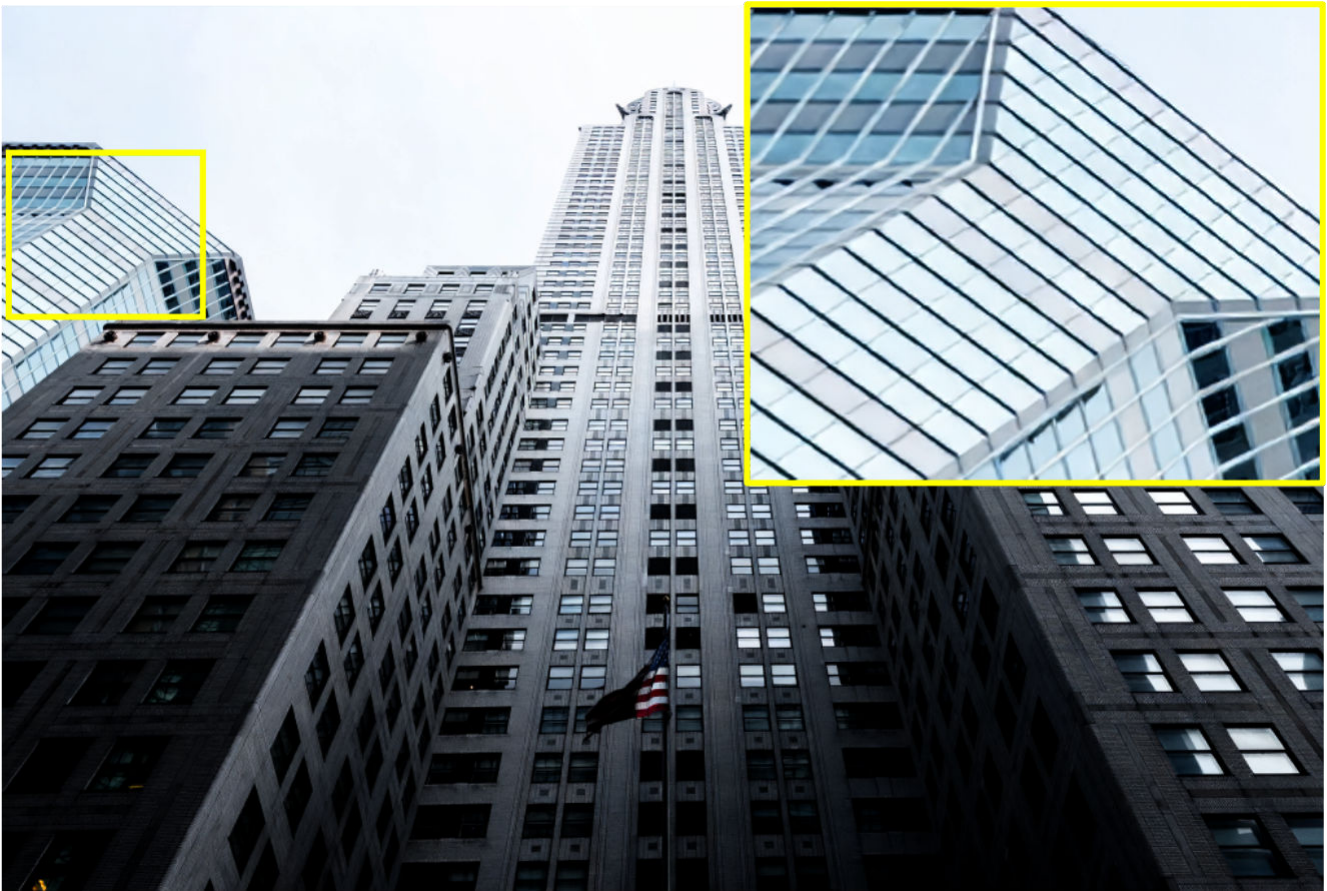} \\
\makebox[0.24\textwidth]{WeConvene (0.137)~\cite{WeConvene_ECCV2024}}
\makebox[0.24\textwidth]{HiFiC (0.157)~\cite{HiFiC_NeurIPS2020}} 
\makebox[0.24\textwidth]{MRIC (0.105)~\cite{MRIC_CVPR2023}}
\makebox[0.24\textwidth]{CDC (0.376)~\cite{CDC_NeurIPS2024}} \\
\includegraphics[width=0.24\textwidth]{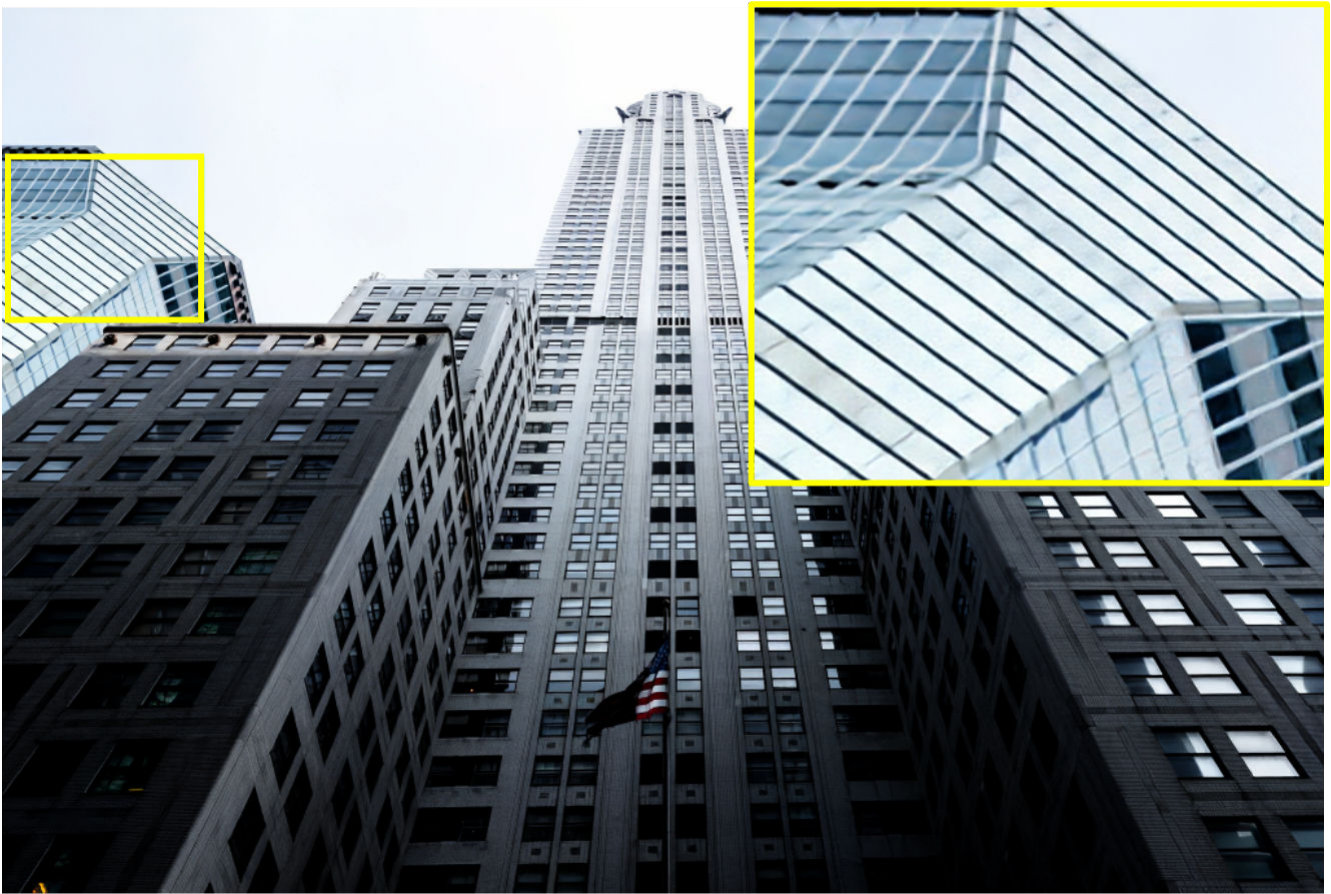} 
\includegraphics[width=0.24\textwidth]{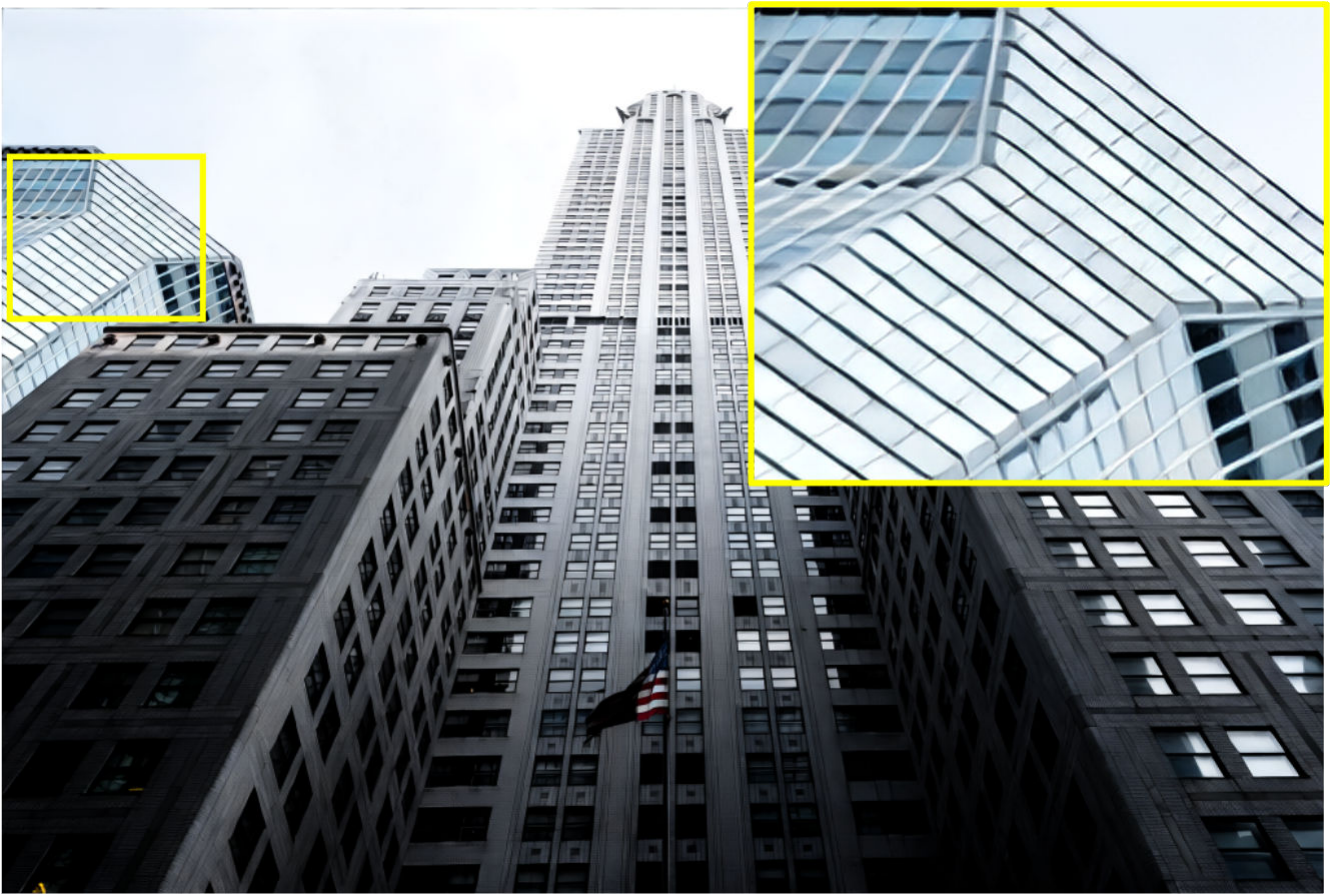} 
\includegraphics[width=0.24\textwidth]{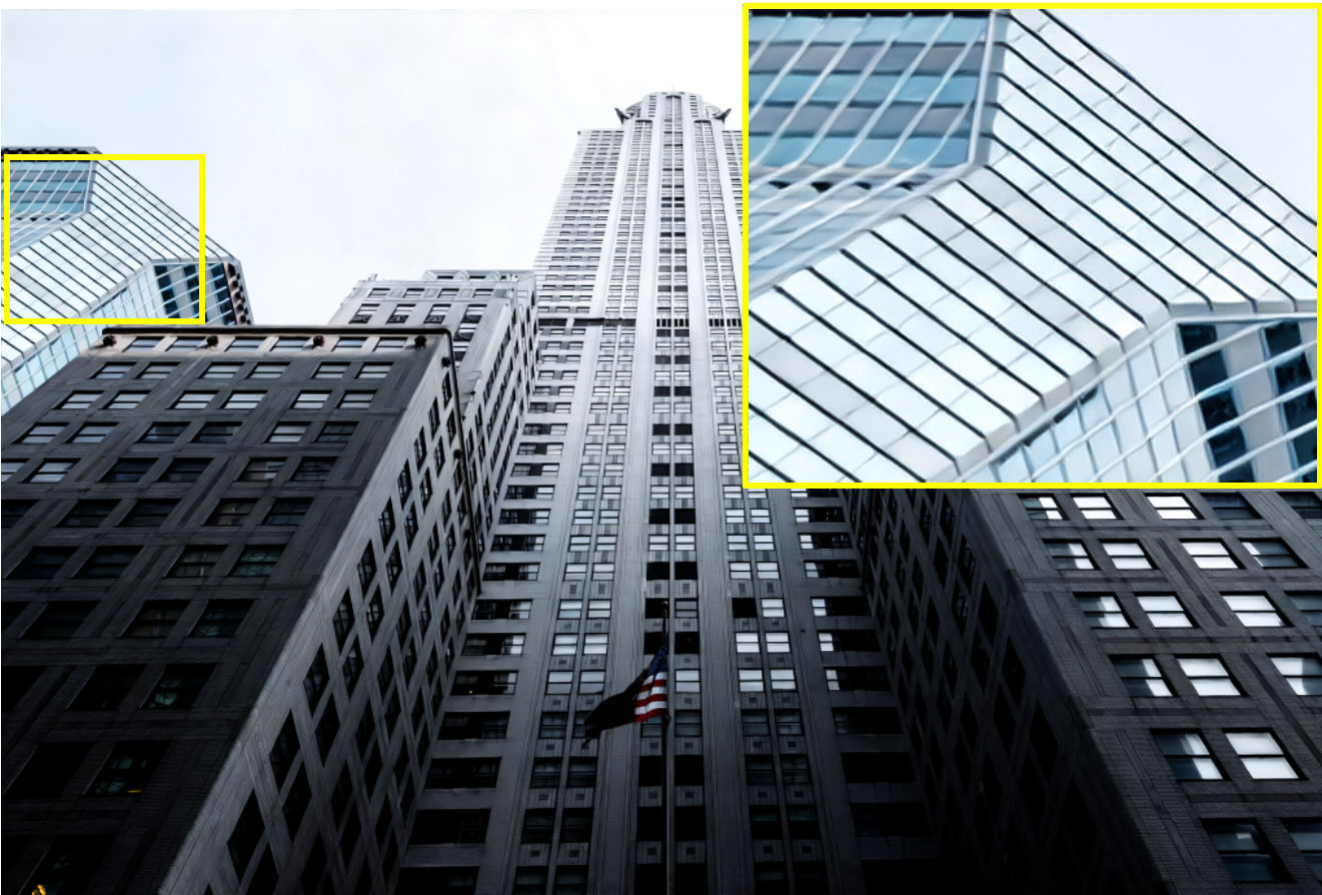} 
\includegraphics[width=0.24\textwidth]{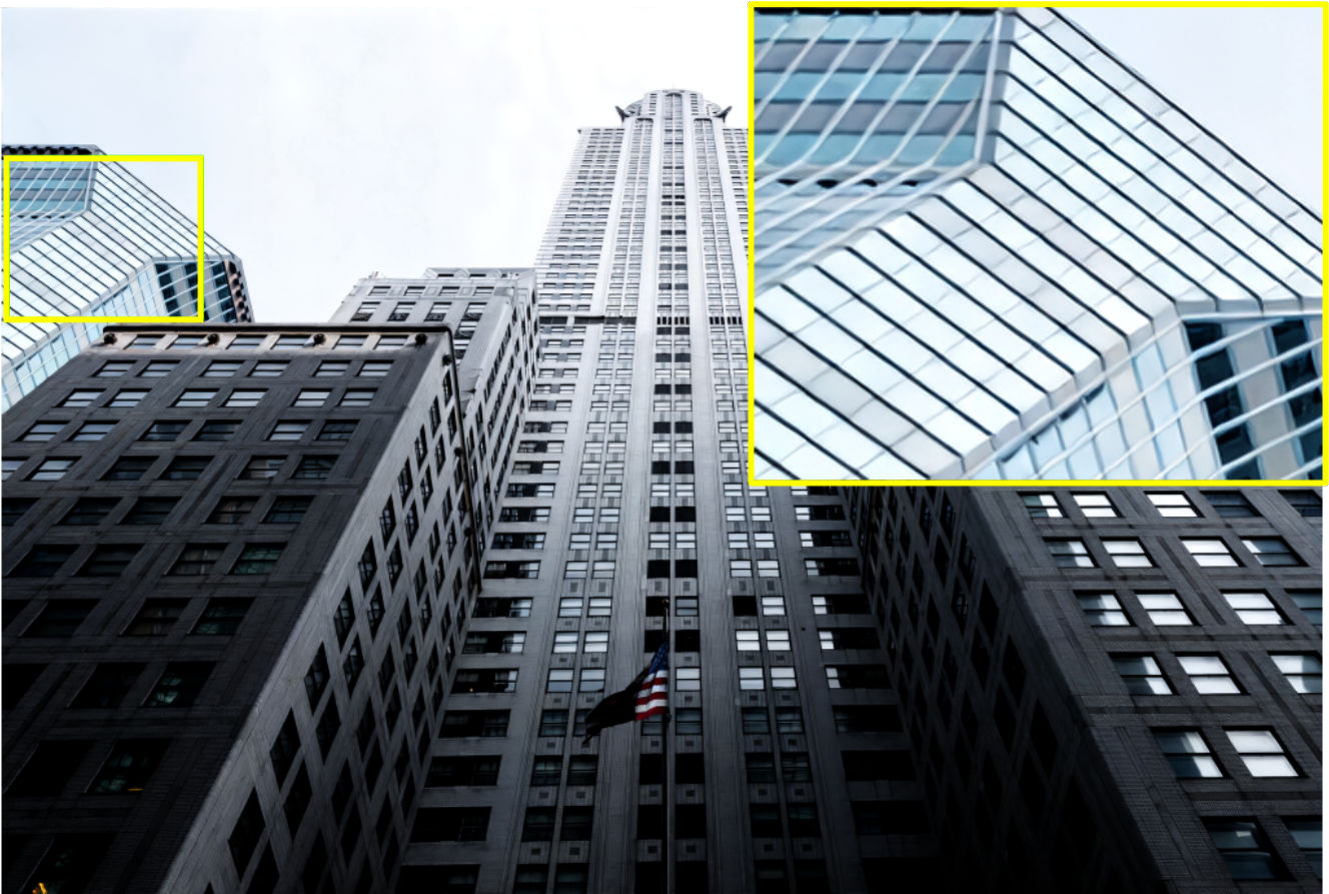} \\
\makebox[0.24\textwidth]{MS-ILLM (0.137)~\cite{MS_ILLM_PMLR2023}}
\makebox[0.24\textwidth]{ICISP (Ours)-$\lambda=5$ (\textbf{0.099})}
\makebox[0.24\textwidth]{ICISP (Ours)-$\lambda=2.5$ (0.158)}
\makebox[0.24\textwidth]{ICISP (Ours)-$\lambda=1.5$ (0.219)}\\
\caption{Visual comparisons on the DIV2K dataset (846.png). Best values are in bold font. Please zoom in for best view.}
\label{vis_com_div2k_846}
\end{figure*}

Fig.~\ref{vis_com_div2k_846} presents the results of the visual comparison. The competing methods produce either overly smooth or unrealistic results, whereas our method delivers more realistic results with texture detail at lower bitrates, successfully preserving the structural details of the building.

\subsubsection{Complexity comparisons} 
We further compare the model complexity of the proposed method and state-of-the-art methods in terms of parameters, FLOPs, and inference time. As shown in Table~\ref{complexity_comparison}, our ICISP has the fewest parameters and the lowest FLOPs. Moreover, although CDC offers the fastest encoding, it requires more time for decoding due to its iterative denoising steps. In contrast, our method achieves the second-best performance against other methods regarding inference time.
\begin{table*}
  \centering
  \caption{Comparison of model complexity, evaluated on the Kodak dataset. The inference time is tested on a machine with a single NVIDIA GeForce RTX 3090 GPU. The BD-rate is computed based on the LPIPS-BPP curve. Best and second-best performances are highlighted in bold and underlined, respectively. The symbol $\dag$ denotes the proposed heavy model.}  
  \setlength{\tabcolsep}{6mm}{
  \begin{tabular}{cccccccc}
    \toprule
    \multirow{2}{*}{Methods} & \multirow{2}{*}{Parameters (M)} & \multirow{2}{*}{FLOPs (G)} & \multicolumn{3}{c}{Inference time (ms)} & \multirow{2}{*}{BD-rate (\%)$\downarrow$}\\
    \cline{4-6} 
    &  &  & Encoding & Decoding & Average  & \\
    \midrule
    WeConvene~\cite{WeConvene_ECCV2024} &  107.15 & 904.56 & 296.67 & 246.79 & 271.73 & 149.78 \\
    TCM~\cite{TCM_CVPR2023} & \underline{45.18} & \underline{211.37} & 163.90 & 135.65 & 149.78 & 156.27 \\
    FTIC~\cite{FAT_ICLR2024} &  70.96 & 245.46 & - & - & - & 136.18 \\ 
    HiFiC~\cite{HiFiC_NeurIPS2020} &  181.57 & 383.48 & 444.24 & 1041.22 & 742.73 & 85.93 \\
    MRIC~\cite{MRIC_CVPR2023} & 89.65 & 820.03 & 2096.72 & 1206.17 & 1651.45 & 89.41 \\
    MS-ILLM~\cite{MS_ILLM_PMLR2023} & 181.48 & 383.48 & \underline{90.11} & \textbf{100.85} & \textbf{95.48} & 51.09 \\
    CDC ~\cite{CDC_NeurIPS2024} & 53.89 & 806.57 & \textbf{32.77} & 4444.91 & 2238.84 & \underline{46.26} \\ 
    TACO~\cite{TACO_ICML2024} & 101.75 & 328.85 & 115.65 & 105.62 & 110.64 & 64.58 \\
    \midrule
    ICISP (Ours) & \textbf{29.26} & \textbf{114.08} & 100.13 & \underline{104.79} & \underline{102.46} & \textbf{0} \\
    \bottomrule
  \end{tabular}}
  \label{complexity_comparison}
\end{table*}

\subsection{Analysis and Discussion}

\begin{table}[htbp]
\centering
\caption{Effectiveness of each component in the proposed compression model, evaluated on the Kodak dataset. The BD-rate is computed on the PSNR-BPP and LPIPS-BPP curve.}
\begin{tabular}{ccccc}
     \toprule
     \multirow{2}{*}{Methods} & \multicolumn{2}{c}{BD-rate (\%)$\downarrow$} & \multirow{2}{*}{Parameters (M)} & \multirow{2}{*}{FLOPs (G)} \\
     \cline{2-3}
     & PSNR & LPIPS  & & \\
     \midrule
    VSSM+FFN & 15.99 & 2.39 & 25.99 & 62.91 \\
    EVSSB+FFN & 9.27 & 2.19 & 27.48 & 105.90 \\
    EVSSB+VSSM & 4.27 & 2.30 & 27.74 & 109.31 \\
    \midrule
    EVSSB+FDMB & 0 & 0 & 29.26 & 114.08 \\
    \bottomrule    
\end{tabular}
\label{ablation_study}
\end{table}

\subsubsection{Effectiveness of EVSSB}
\begin{figure}
\centering
\includegraphics[width=0.155\textwidth]{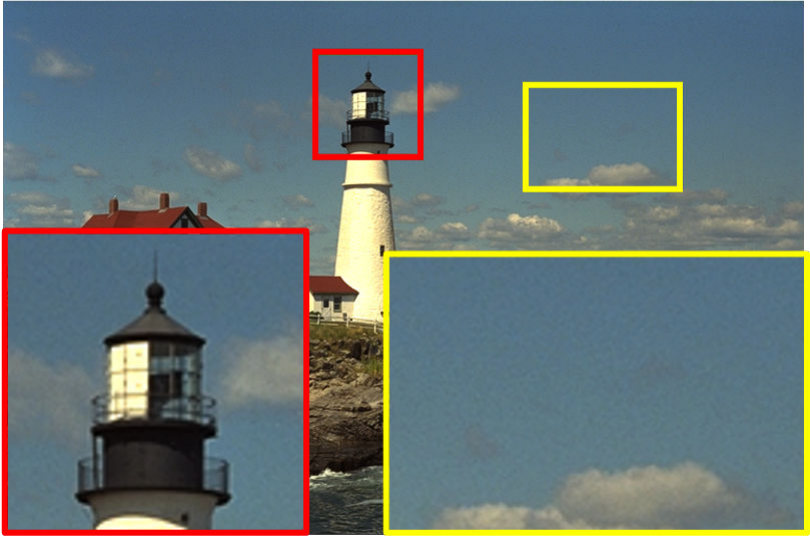} 
\includegraphics[width=0.155\textwidth]{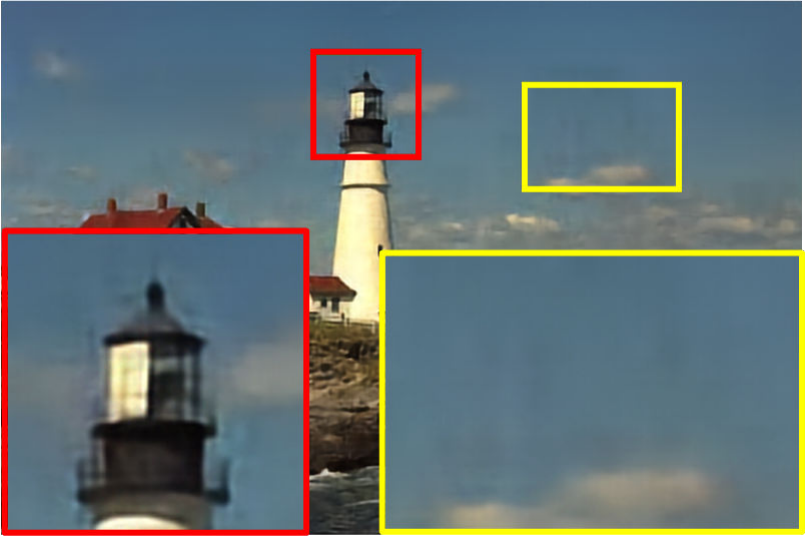} 
\includegraphics[width=0.155\textwidth]{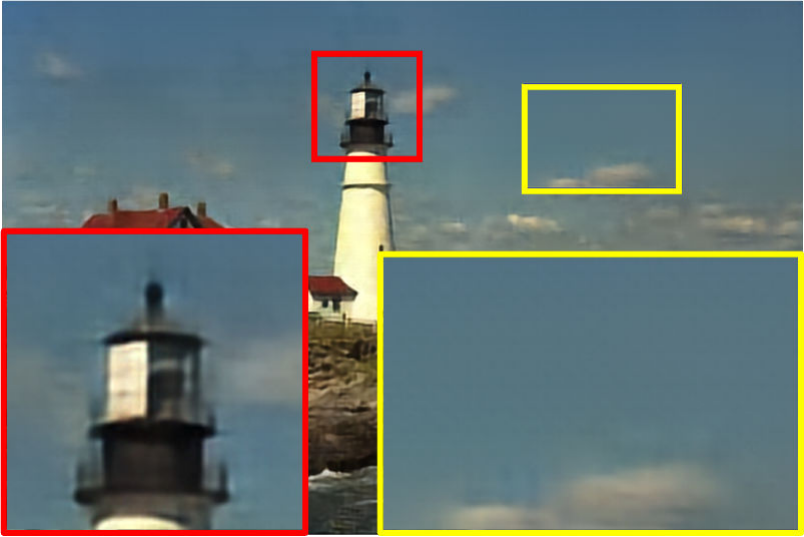} \\
\makebox[0.15\textwidth]{(a) Original image}
\makebox[0.15\textwidth]{(b) w/o EVSSB}
\makebox[0.15\textwidth]{(c) w/ EVSSB}
\caption{Effectiveness of EVSSB. The [LPIPS@Bitrates] of (b) and (c) are [0.4280@0.1363bpp] and [0.4230@0.1345bpp]. Zoom in for best view.}
\label{EVSSB_ablation}
\end{figure}
The proposed EVSSB is used to explore the local and non-local spatial relationships for redundancy reduction. 
To demonstrate the effectiveness of EVSSB, we remove EVSSB from the proposed method and use VSSM by default. We train this baseline using the same experimental setting as ours for a fair comparison. 

Table~\ref{ablation_study} shows the quantitative results on the Kodak dataset. Using the EVSSB produces better results with smaller BD-rate value~\cite{BD_Rate_2001}, achieving 6.72\% and 0.2\% rate reduction in terms of PSNR and LPIPS, respectively, compared to the method without using the EVSSB (see comparisons of ``VSSM+FFN" and ``EVSSB+FFN"). The visual comparisons in Fig.~\ref{EVSSB_ablation}(b) and (c) further demonstrate that using the proposed EVSSB facilitates image compression, where the lighthouse and sky are well reconstructed with fewer artifacts.

\subsubsection{Effect of FDMB}
The proposed FDMB separates the low-frequency and high-frequency features, which are gated by separate VSSMs to adaptively decide whether to preserve or eliminate information. 
To evaluate whether FDMB can contribute to image compression, we replace the FDMB with the vanilla feed-forward network (FFN) and retrain the baseline model from scratch. Table~\ref{ablation_study} shows that using the FDMB generates better results in terms of BD-rate, where the values are 9.27\% (PSNR) and 2.19\% (LPIPS) lower than the method using the FDMB (``EVSSB+FFN" vs. ``EVSSB+FDMB") .
\begin{figure}[htbp]
\centering
\includegraphics[width=0.155\textwidth]{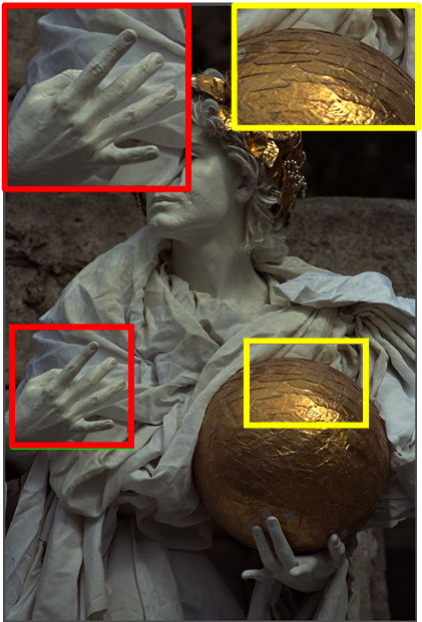} 
\includegraphics[width=0.155\textwidth]{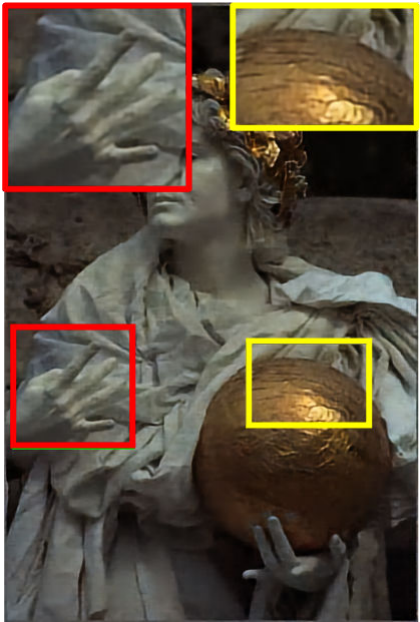} 
\includegraphics[width=0.155\textwidth]{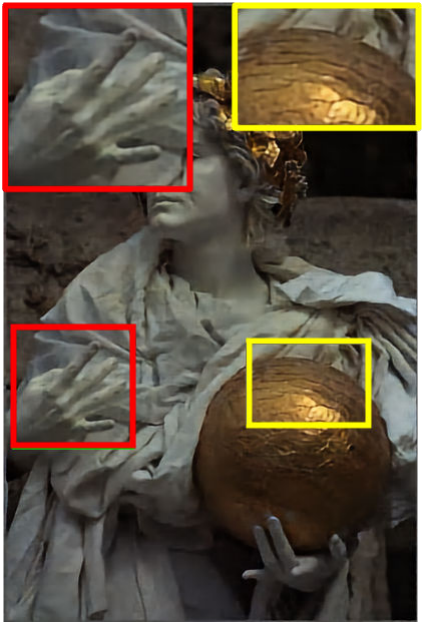} \\
\makebox[0.15\textwidth]{(a) Original image}
\makebox[0.15\textwidth]{(b) w/o FDMB}
\makebox[0.15\textwidth]{(c) w/ FDMB}
\caption{Effectiveness of FDMB. The [LPIPS@Bitrates] of (b) and (c) are [0.3856@0.1573bpp] and [0.3750@0.1514bpp]. Zoom in for best view.}
\label{FDMB_ablation}
\end{figure}

In addition, we evaluate the effect of frequency decomposition in the FDMB. For this purpose, we further remove the Haar transform and its inverse version and use only the VSSM. As shown in Table~\ref{ablation_study}, utilizing the frequency decomposition can achieve better compression results, where the BD-rate is 4.27\% lower than the method without frequency decomposition (see comparisons of ``EVSSB+VSSM" and ``EVSSB+FDMB"). Fig.~\ref{FDMB_ablation} also demonstrates that using the FDMB can yield better reconstruction results; for example, the hand of the sculpture and the ball are well restored at lower bitrates.
\subsubsection{Impact of implicit priors} 
\begin{figure}
\centering
\includegraphics[width=0.24\textwidth]{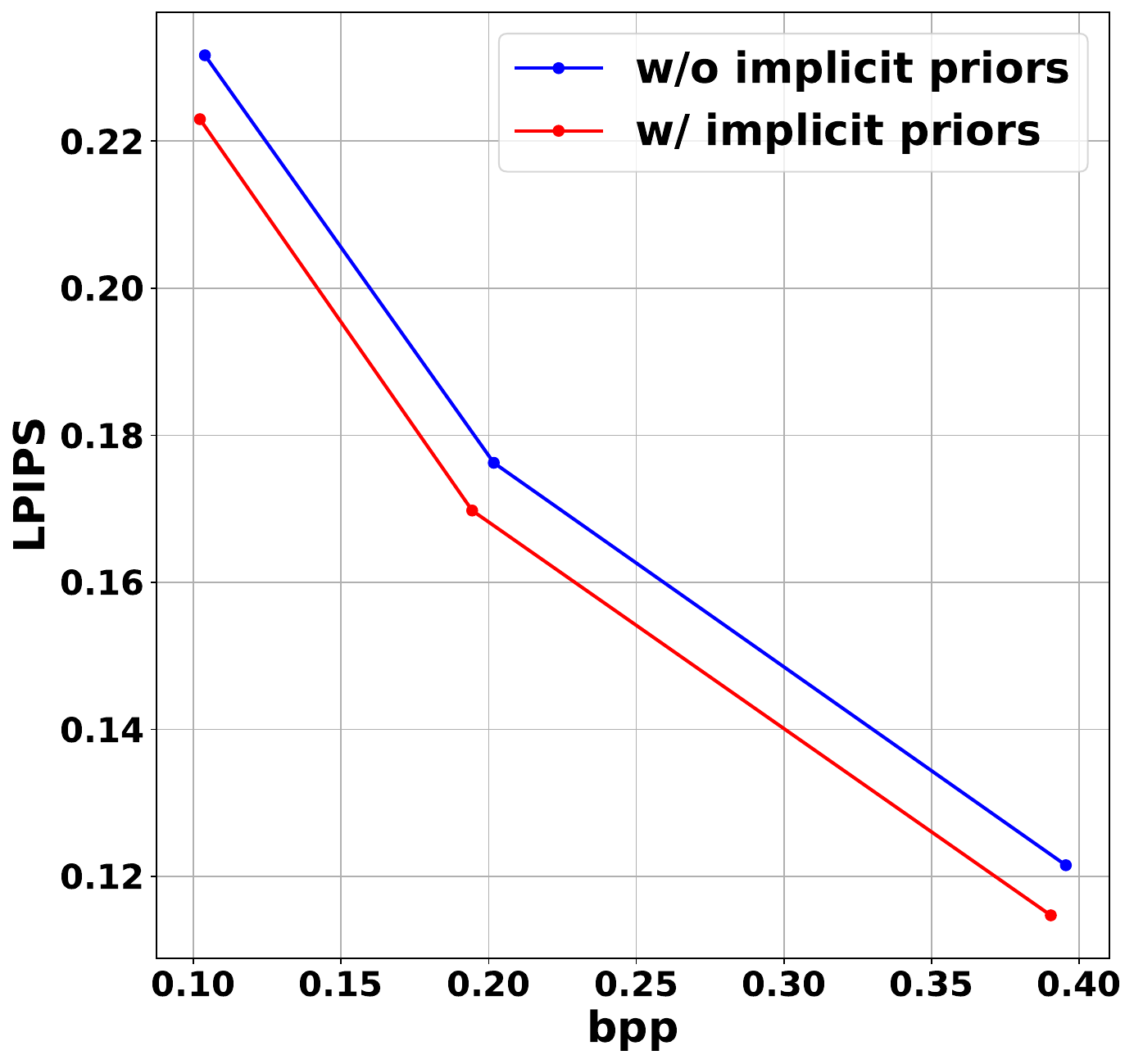} 
\includegraphics[width=0.24\textwidth]{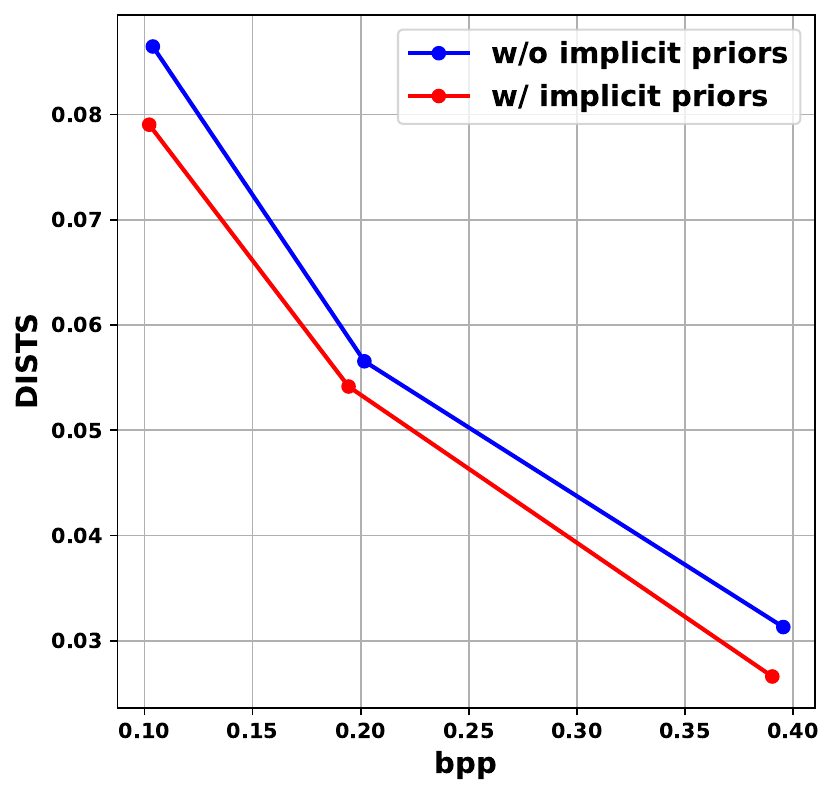}
\caption{Ablation studies of the implicit priors. Lower LPIPS/DISTS indicate better performance.}
\label{IP_ablation_quan}
\end{figure}
\begin{figure}
\centering
\includegraphics[width=0.155\textwidth]{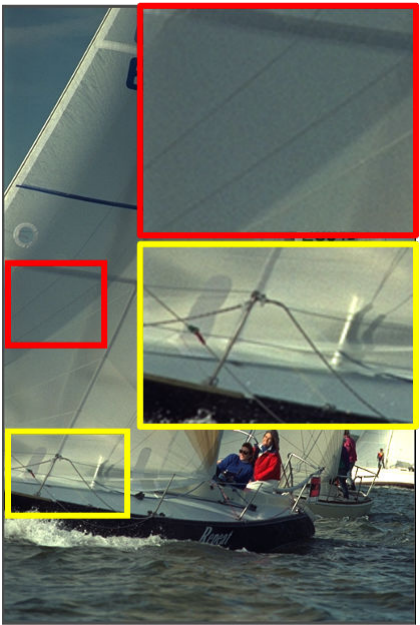} 
\includegraphics[width=0.155\textwidth]{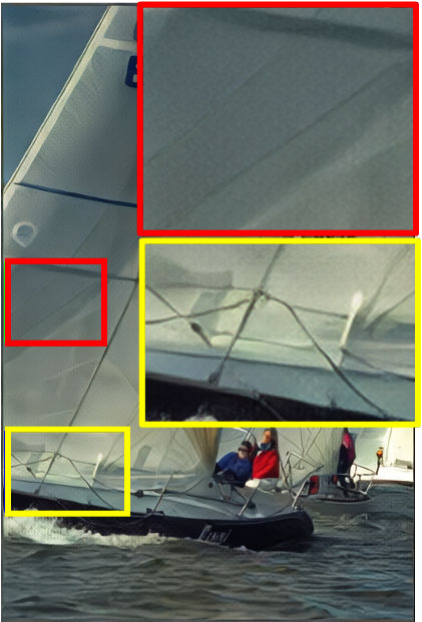} 
\includegraphics[width=0.155\textwidth]{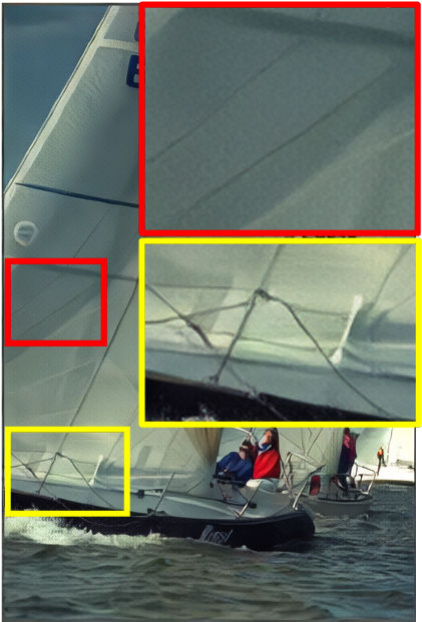} \\
\makebox[0.15\textwidth]{(a) Original image}
\makebox[0.15\textwidth]{(b) w/o IP}
\makebox[0.15\textwidth]{(c) w/ IP}\\
\caption{Effectiveness of implicit priors (IP for short). The [LPIPS@Bitrates] of (b) and (c) are [0.2121@0.1033bpp] and [0.2004@0.1008bpp]. Zoom in for best view.}
\label{IP_ablation}
\end{figure}

Our goal is to exploit the implicit semantic priors in the pretrained DINOv2 model to facilitate the semantic texture generation of the compression model at low bitrates. To verify its effectiveness, we remove the implicit priors and retrain the proposed method. As shown in Fig.~\ref{IP_ablation_quan}, we find that using the implicit priors can bring a performance gain, especially at the low bitrates. Fig.~\ref{IP_ablation} also indicates that using the implicit priors generates better visual results, where the structure of the sails and boat rails are reconstructed at lower bitrates. 

\subsubsection{Effectiveness of AFFB and DSFT}
\begin{figure}
\centering
\includegraphics[width=0.24\textwidth]{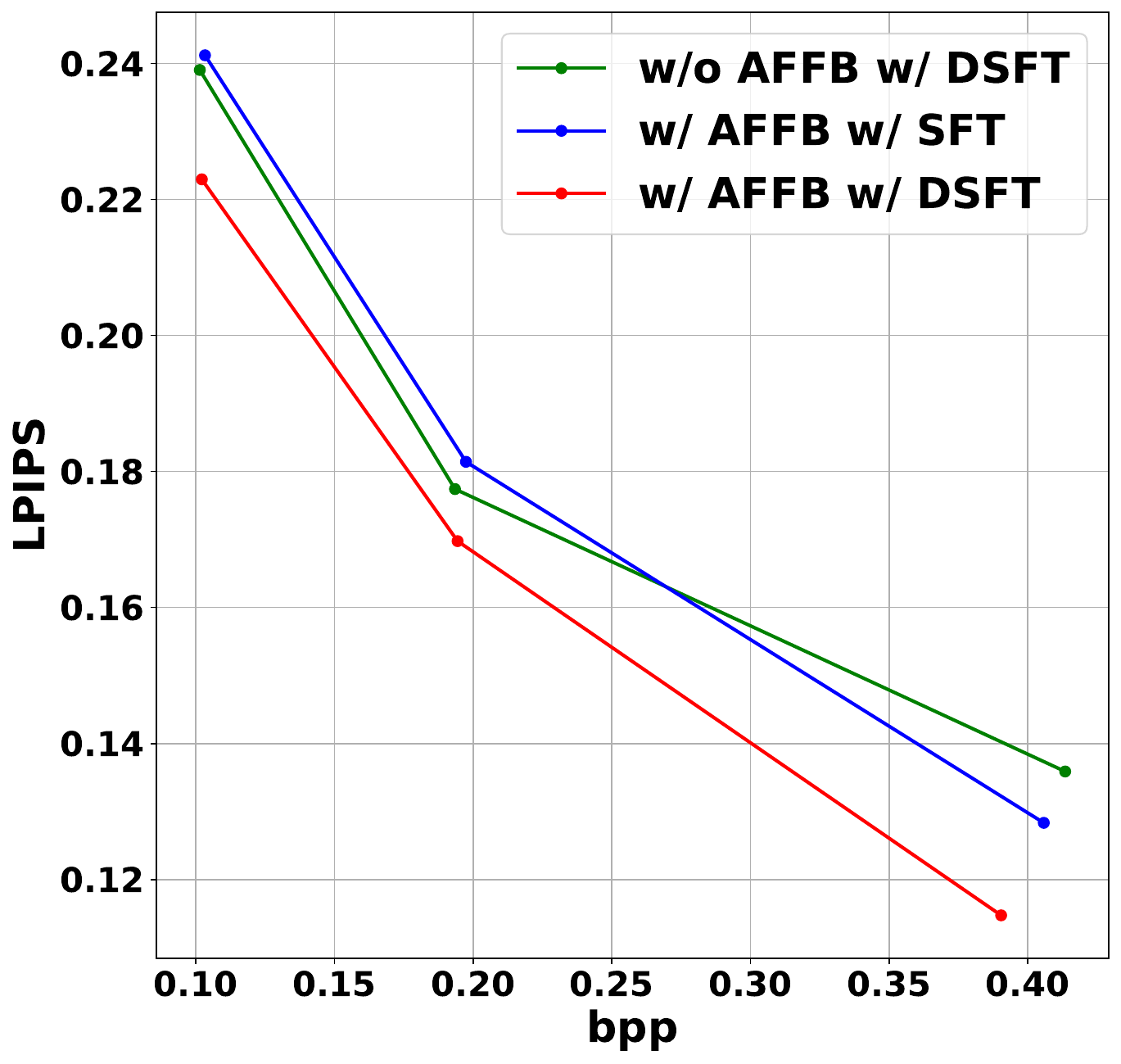} 
\includegraphics[width=0.24\textwidth]{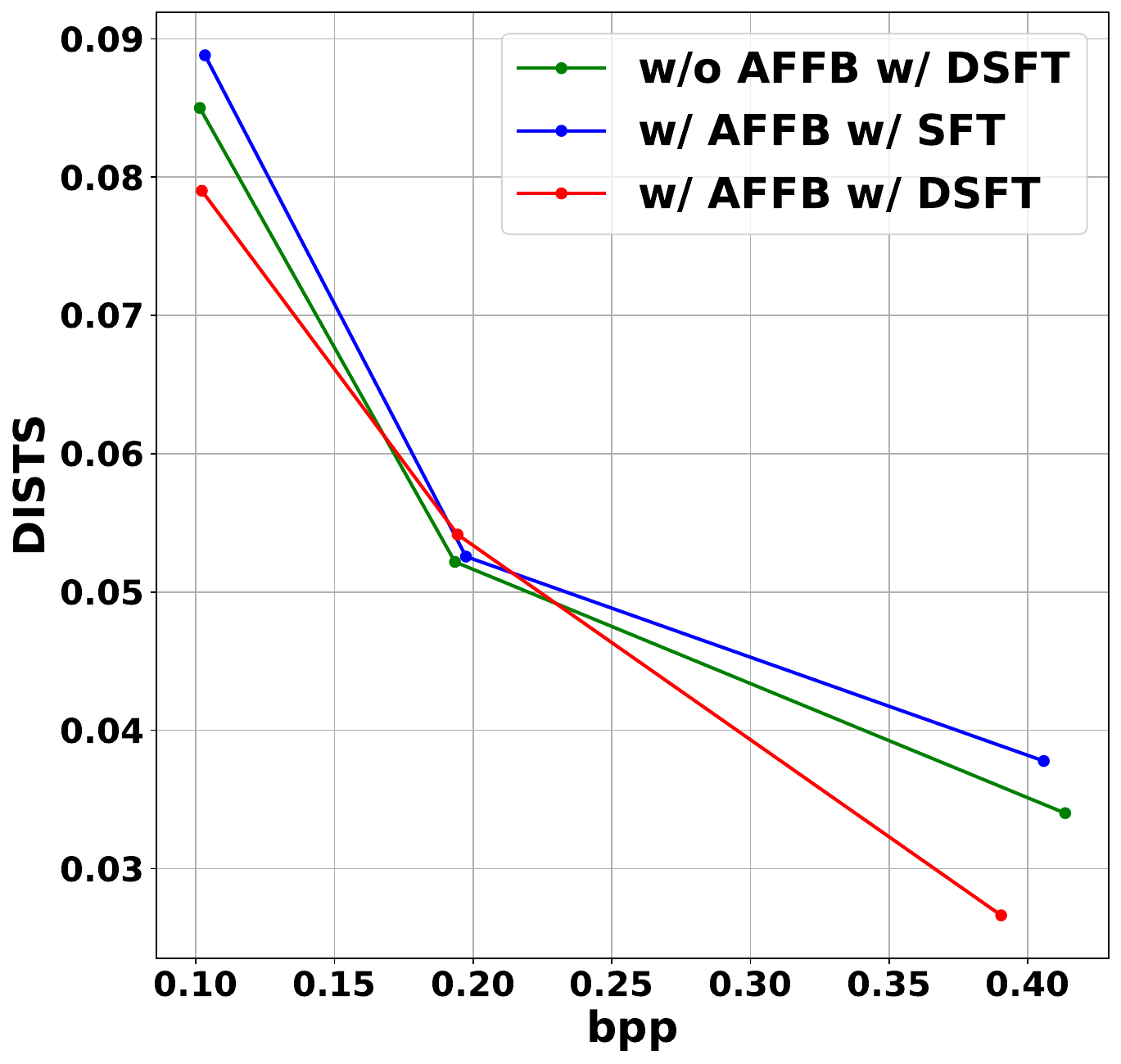}
\caption{Effectiveness of AFFB and DSFT.}
\label{effectiveness_AFFB_DSFT}
\end{figure}
The proposed AFFB aims to fuse the refined latent $\mathbf{\bar{y}}$ and semantic features $\mathbf{F}_{sp}$ for effective condition generation. To validate its effectiveness, we use the concatenation operation followed by convolutions instead and train the proposed method using the same experimental settings for a fair comparison. As shown in Fig.~\ref{effectiveness_AFFB_DSFT}, we find that using AFFB can bring performance gains, where the LPIPS/DISTS values are smaller than the method without using AFFB (see red curves vs. green curves)

In addition, we further validate the effectiveness of DSFT, which is proposed to effectively incorporate the condition into the proposed semantic-informed discriminator. We use the SFT used in~\cite{GFPGAN_CVPR2021} by default. Fig.~\ref{effectiveness_AFFB_DSFT} demonstrates that using DSFT results in performance improvements, as evidenced by the smaller LPIPS/DISTS values compared to the method with SFT (red curves vs. blue curves).

\subsubsection{Visualization features of semantic-informed discriminator}
\begin{figure}
\centering
\includegraphics[width=0.23\textwidth]{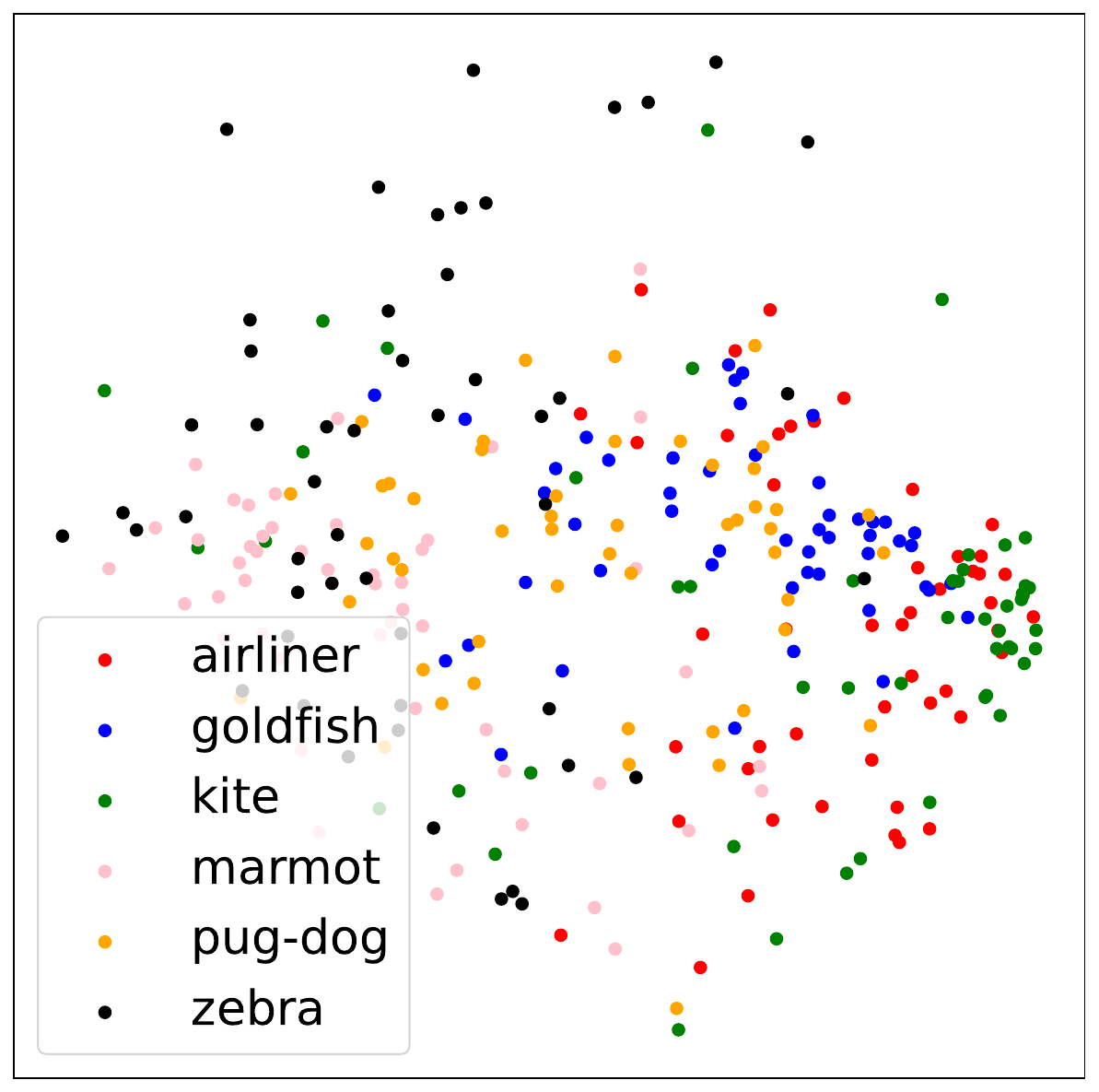} 
\includegraphics[width=0.23\textwidth]{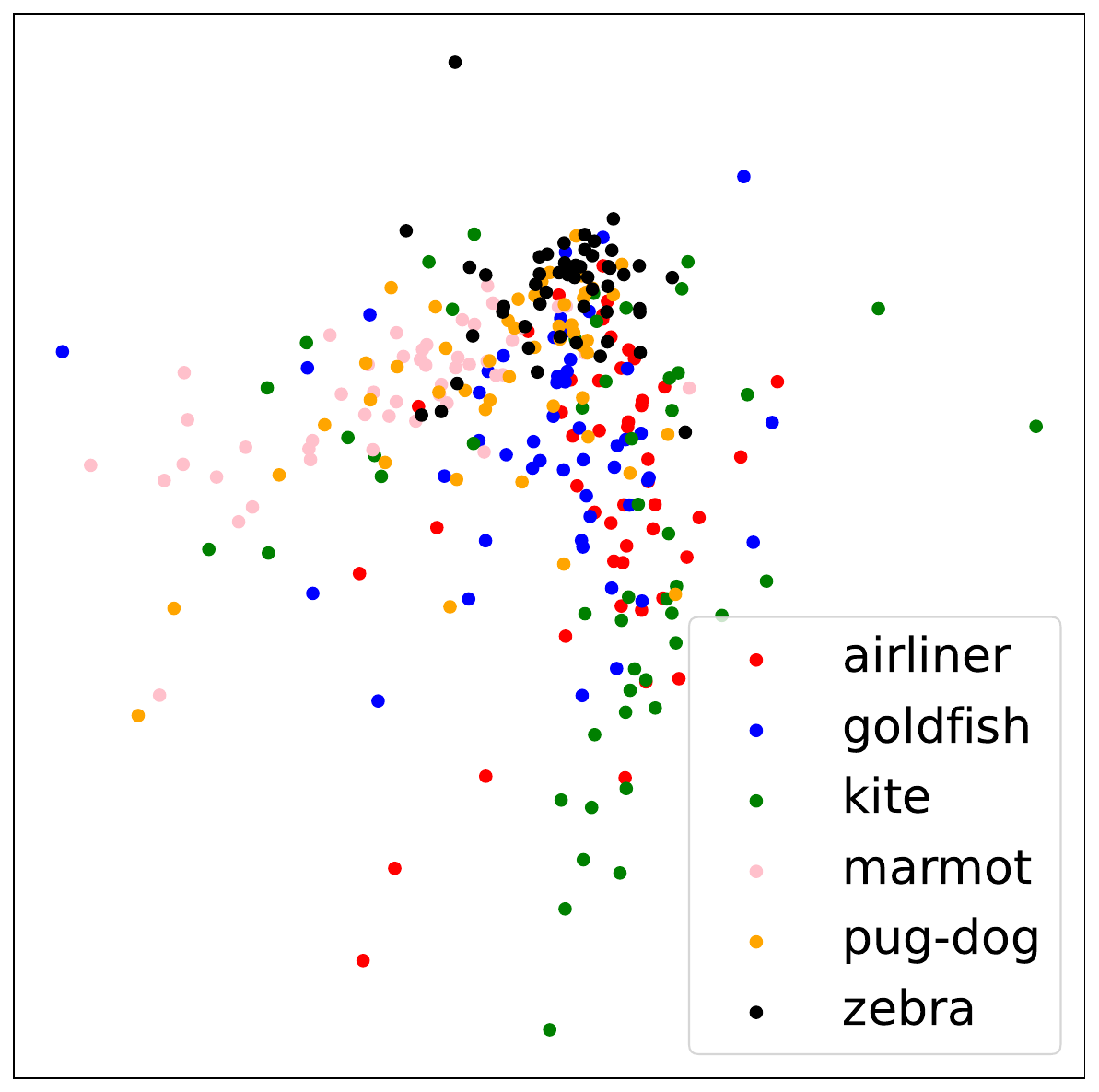} 
\makebox[0.2\textwidth]{(a) w/o implicit priors}
\makebox[0.2\textwidth]{(b) w/ implicit priors}
\caption{The t-SNE~\cite{tsne} visualization of features in the discriminator without and with implicit priors.}
\label{vis_dis}
\end{figure}
To verify whether the proposed semantic-informed discriminator is able to discriminate semantics, we use the t-SNE~\cite{tsne} to visualize the features of the discriminator. Specifically, we select six classes from ImageNet~\cite{ImageNet22K}, where each class has 50 images, and feed them into the discriminator without and with implicit priors. As shown in Fig.~\ref{vis_dis}, we find that using the implicit priors can help the semantic features to cluster better. This indicates that the proposed semantic-informed discriminator can show better discriminative ability in different semantics, thus facilitating the compression model to generate semantic texture details at low bitrates. 
\subsubsection{Comparison with standard nonlinear transforms}
\begin{table}[htbp]
  \centering
  \caption{Comparison with other nonlinear transforms, evaluated on the Kodak dataset. The BD-rate is computed on the PSNR-BPP and LPIPS-BPP curve.}  
  \begin{tabular}{ccccc}
     \toprule
     \multirow{2}{*}{Methods} & \multicolumn{2}{c}{BD-rate (\%)$\downarrow$} & \multirow{2}{*}{Parameters (M)} & \multirow{2}{*}{FLOPs (G)} \\
     \cline{2-3}
     & PSNR & LPIPS  & & \\
     \midrule
    CNN-based & 80.36 & 9.28  & 29.43 & 150.36 \\
    SwinT-based & 39.90 & 7.36 & 29.41 & 115.81 \\
    Mamba-based & 29.45 & 19.99 & 29.28 & 66.44 \\
    \midrule
    RSSM-based & 0 & 0 & 29.26 & 114.08 \\
    \bottomrule    
\end{tabular}
\label{transforms_com}
\end{table}

\begin{figure}
\centering
\includegraphics[width=0.24\textwidth]{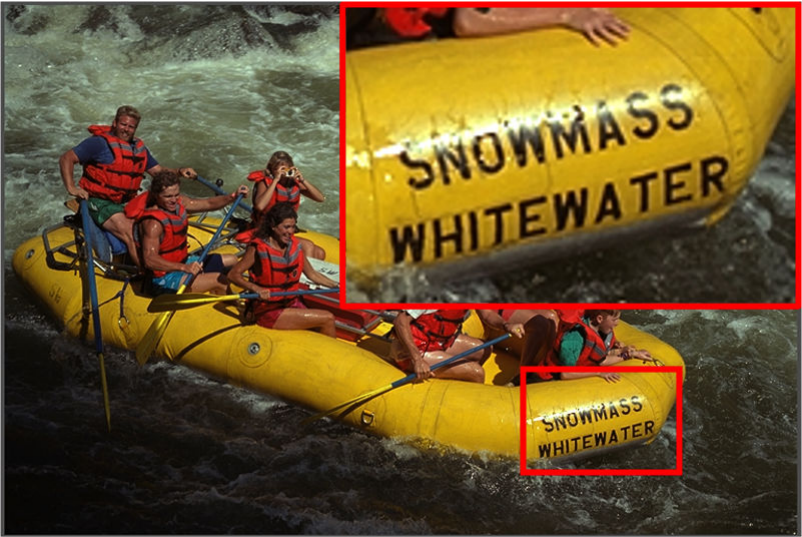}\\
\makebox[0.24\textwidth]{(a) Original image}\\
\includegraphics[width=0.24\textwidth]{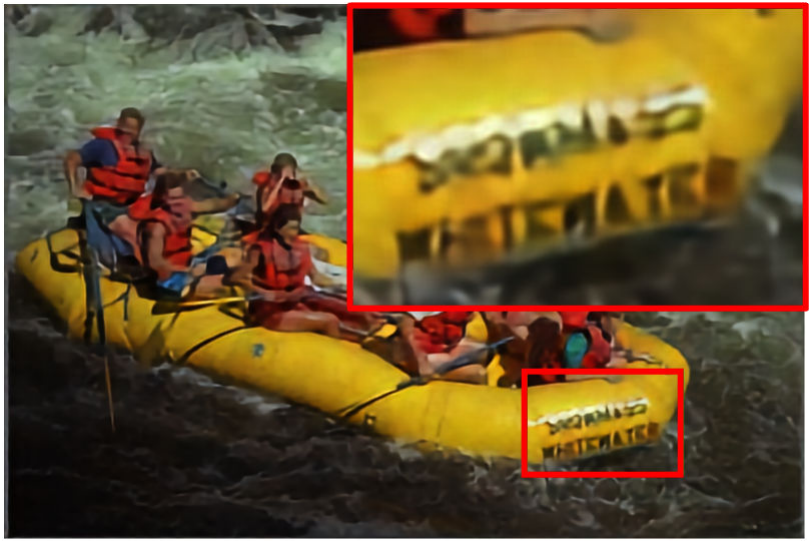} 
\includegraphics[width=0.24\textwidth]{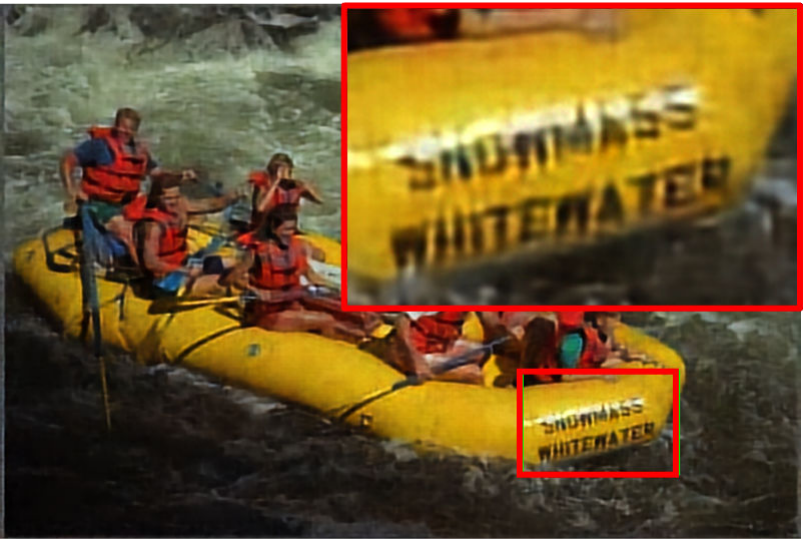}\\
\makebox[0.24\textwidth]{(b) CNN-based \cite{hyperprior}}
\makebox[0.24\textwidth]{(c) SwinT-based \cite{TTC_ICML2021}}\\
\includegraphics[width=0.24\textwidth]{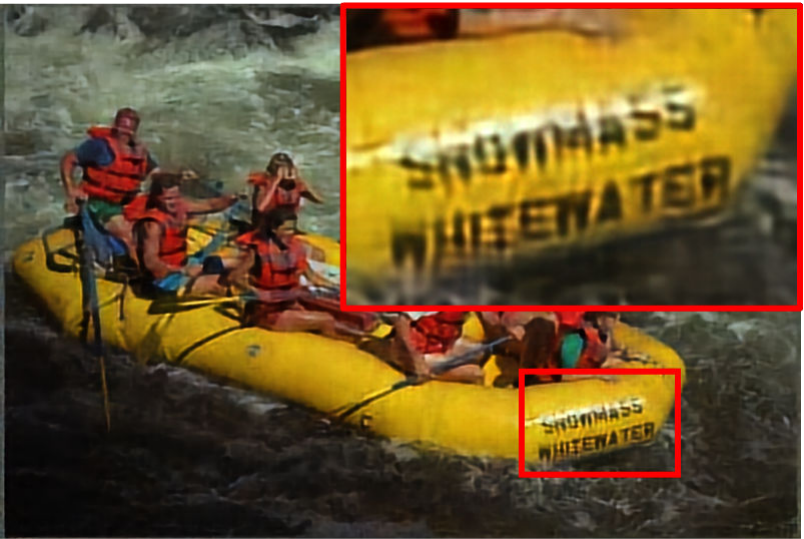} 
\includegraphics[width=0.24\textwidth]{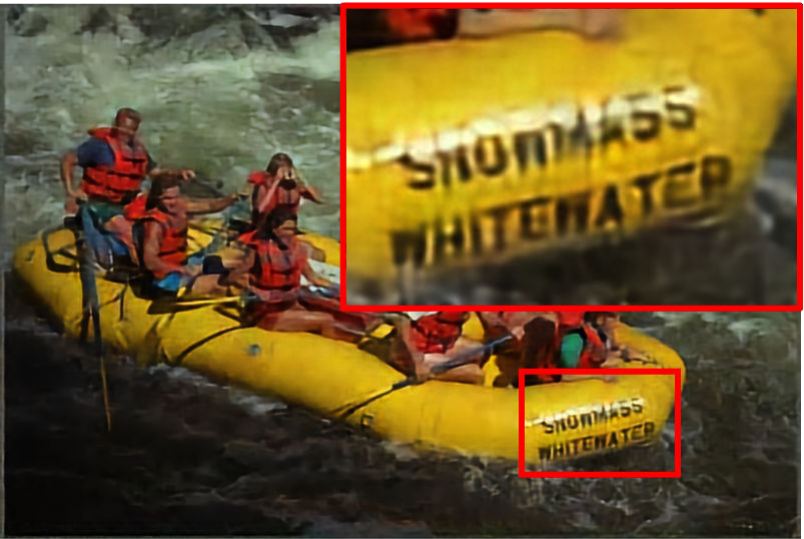} \\
\makebox[0.24\textwidth]{(d) Mamba-based \cite{MambaVC_2024}}
\makebox[0.24\textwidth]{(e) RSSM-based (proposed)}\\
\caption{Visual comparisons of different nonlinear transforms on the Kodak dataset. The [LPIPS@Bitrates] of (b)-(e) are [0.4319@0.1964bpp], [0.4155@0.1980bpp], [0.4070@0.2053bpp] and [0.4107@0.1775bpp]. Zoom in for best view.}
\label{transforms_vis_com}
\end{figure}

We compare the proposed RSSM-based transform with three representative nonlinear transforms, including the CNN-based~\cite{hyperprior}, the Swin Transformer-based~\cite{TTC_ICML2021} (SwinT for short), and the Mamba-based transform~\cite{MambaVC_2024}. For fair comparisons, we make these methods have the same network parameters as our method. As shown in Table~\ref{transforms_com}, our method achieves the bitrate savings of 80.36\%, 39.90\%, and 29.45\% compared to the CNN-based, SwinT-based, and Mamba-based transforms in terms of PSNR, respectively. Fig.~\ref{transforms_vis_com} also shows that the RSSM-based transform can produce better results at lower bitrates with clearer characters.

\subsubsection{Robustness to different pretrained vision models}
\begin{figure}
\centering
\includegraphics[width=0.24\textwidth]{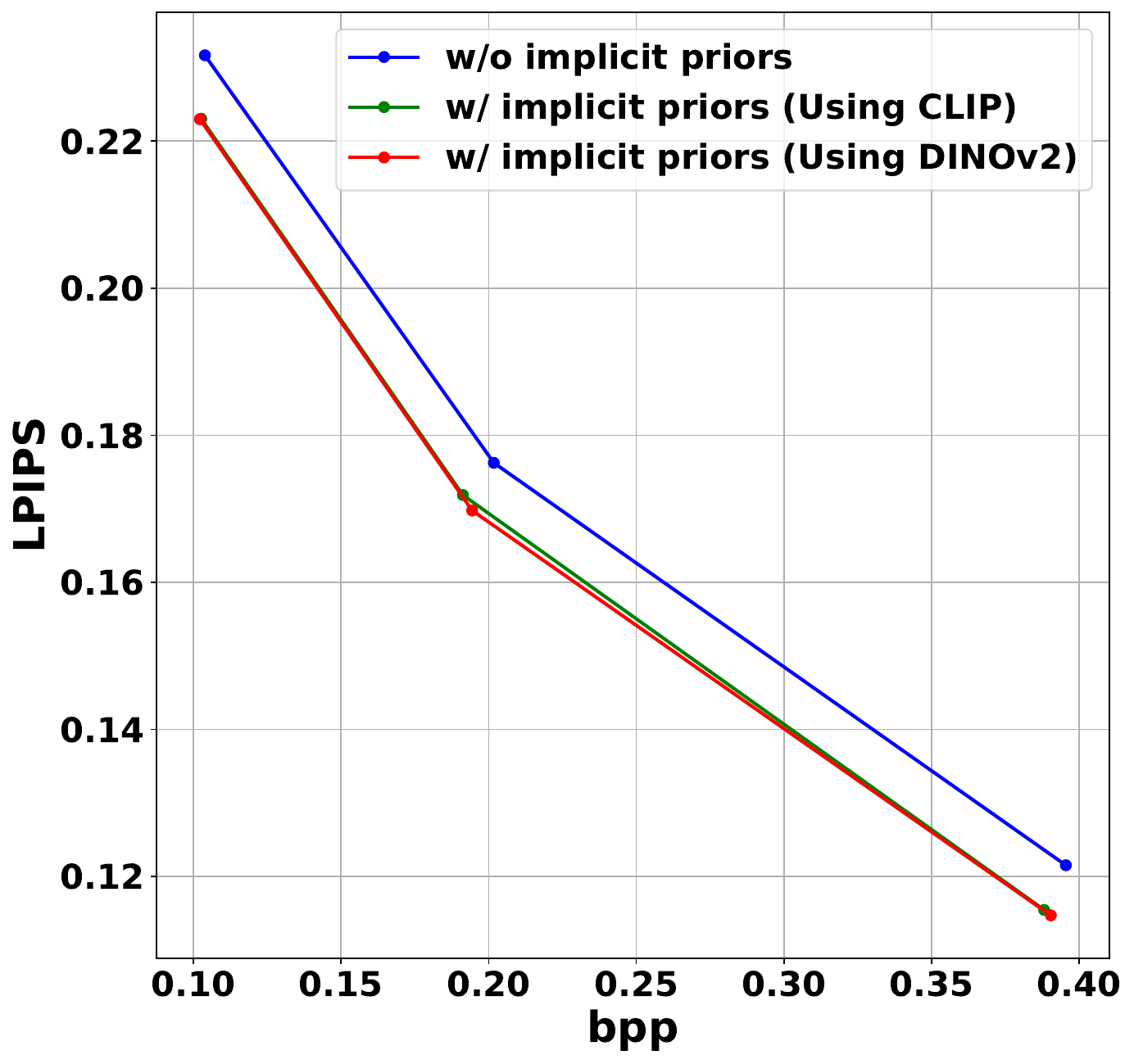} 
\includegraphics[width=0.24\textwidth]{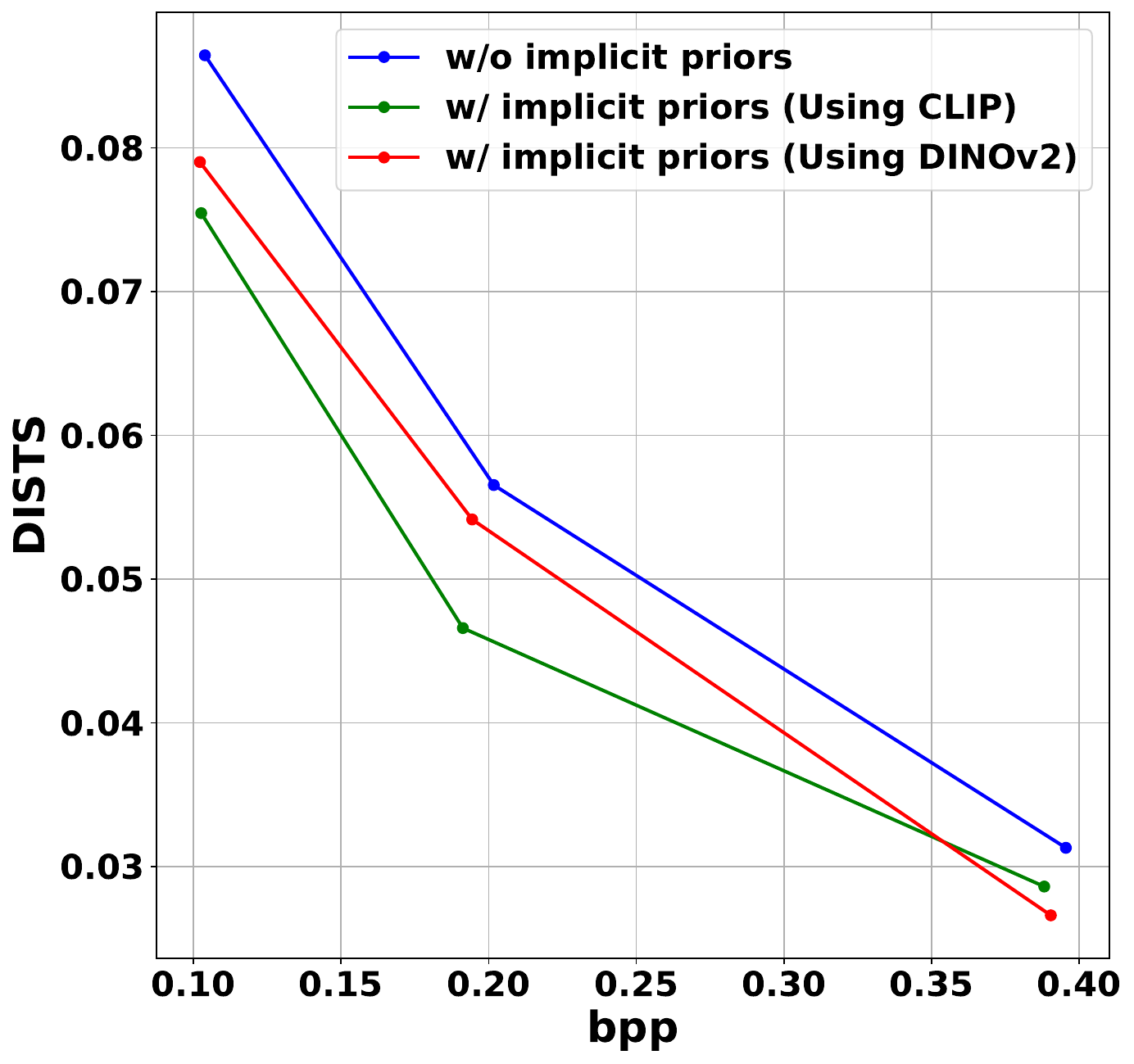}
\caption{Robustness to different pretrained vision models for implicit semantic priors generation. Lower LPIPS/DISTS indicate better performance.}
\label{dino_clip_ablation_quan}
\end{figure}
To test the robustness of the proposed semantic-informed discriminator, we also resort to the large vision-language model CLIP \cite{CLIP}, which has powerful representation capabilities suitable for semantic extraction. Specifically, we replace the DINOv2 encoder with the pre-trained CLIP model and retrain the proposed method under experimental conditions. As shown in Fig. \ref{dino_clip_ablation_quan}, we find that using the CLIP model for implicit semantic priors generation is able to achieve promising results, which further demonstrates the flexibility and robustness of the proposed method.

\subsubsection{Model complexity comparison on images with different resolutions}
We compare the proposed ICISP with other perceptual image compression methods regarding model complexity using images with different resolutions. As shown in Table~\ref{complexity_comparison_2}, our method achieves the lowest FLOPs compared to other approaches, mainly when the image resolution is 2048$\times$2048. In this case, our method reduces FLOPs by 80.79\% compared to TACO~\cite{TACO_ICML2024}.

Moreover, in terms of inference time, our method demonstrates the second-best performance evaluated on low-resolution images and comparable performance on both 1024$\times$1024 and 2048$\times$2048 resolutions. We attribute this to the fact that the 2D selective scan module in the EVSSB and FDMB is time-consuming due to its scanning operation along the four directions~\cite{VMamba_2024}. Future work will focus on optimizing this module to reduce inference time.

\begin{table*}
  \centering
  \caption{Comparison of model complexity, evaluated on the images with different resolutions. The inference time is tested on a machine with a single NVIDIA GeForce RTX 3090 GPU. Best and second-best performances are highlighted in \textbf{bold} and \underline{underlined}, respectively.}
  \setlength{\tabcolsep}{6mm}{
  \begin{tabular}{ccccccc}
    \toprule
    Resolutions  & HiFiC~\cite{HiFiC_NeurIPS2020} & MRIC~\cite{MRIC_CVPR2023} & MS-ILLM~\cite{MS_ILLM_PMLR2023} & CDC~\cite{CDC_NeurIPS2024} & TACO~\cite{TACO_ICML2024} & ICISP (Ours) \\
    \midrule
    \multicolumn{7}{c}{FLOPs (G)}  \\
    \cline{2-7}
    128$\times$128 & 13.85 & 34.18 & 15.98 & 33.61 & \underline{13.65} & \textbf{2.62}\\
    256$\times$256 & 55.42 & 136.68 & 63.91 & 134.43 & \underline{54.58} & \textbf{10.48} \\
    512$\times$512 & 221.68 & 546.69 & 255.65 & 537.72 & \underline{218.28} & \textbf{41.94} \\
    1024$\times$1024 & 886.72 & 2,186.73 & 1,022.61 & 2,150.86 & \underline{873.10} & \textbf{167.75} \\
    2048$\times$2048 & 3,546.87 & 8,746.89 & 4,090.45 & 8,603.46 & \underline{3,492.38} & \textbf{671.00} \\
    \midrule
    \multicolumn{7}{c}{Encoding time (ms)}\\
    \cline{2-7}
    128$\times$128 & \underline{29.68} & 1,845.77 & \textbf{23.36} & 211.71 & 48.25 & 42.38 \\
    256$\times$256 & 82.08 & 1,889.99 & \textbf{29.40} & 211.93 & 53.39 & \underline{47.70} \\
    512$\times$512 & 281.85 & 1,983.62 & \textbf{52.89}  & 293.86 & 76.25 & \underline{68.13} \\
    1024$\times$1024 & 1,076.85 & 2,493.21 & \textbf{145.59} & 300.35 & \underline{182.95} & 201.33 \\
    2048$\times$2048 & 4,089.71 & 5,053.60 & 872.64 & \textbf{302.34} & \underline{662.26} & 780.64 \\
    \midrule
    \multicolumn{7}{c}{Decoding time (ms)}\\
    \cline{2-7}
    128$\times$128 & 50.63 & 856.54 & \textbf{21.25} & 1,059.19 & 41.36 & \underline{36.75} \\
    256$\times$256 & 170.76 & 911.74 & \textbf{31.22} & 1,094.92 & 47.68 & \underline{42.10} \\
    512$\times$512 & 651.79 & 1,053.85 & \textbf{64.02}  & 3,145,16 & 72.97 & \underline{70.37} \\
    1024$\times$1024 & 2,610.29 & 1,640.25 & \underline{190.26} & 11,192.45 & \textbf{174.88} & 205.49 \\
    2048$\times$2048 & 9,859.92 & 3,965.30 & 1,053.96 & 43,823.44 & \textbf{606.28} & \underline{792.60} \\    
    \bottomrule
  \end{tabular}}
  \label{complexity_comparison_2}
\end{table*}

%% file: 06_conclusion.tex
\section{Conclusion}
\label{conclusion}
In this paper, we proposed a lightweight perceptual image compression method using implicit semantic priors.
We developed an enhanced visual state space block to fully capture the local and global dependencies and proposed a frequency decomposition modulation block to adaptively decide which low-frequency and high-frequency information to preserve or eliminate independently. The above two components help the compression network to achieve a compact feature representation.  
Furthermore, we proposed a semantic-informed discriminator that uses the implicit semantic priors from a pretrained DINOv2 encoder, which helps the compression network to generate semantic texture details at low bitrates. 
%
Experimental results demonstrate that the proposed method, with lower model complexity, achieves performance that is either better or comparable to state-of-the-art approaches.

%% file: 09_biography.tex
\begin{IEEEbiography}[{\includegraphics[width=1in,height=1.25in,clip,keepaspectratio]{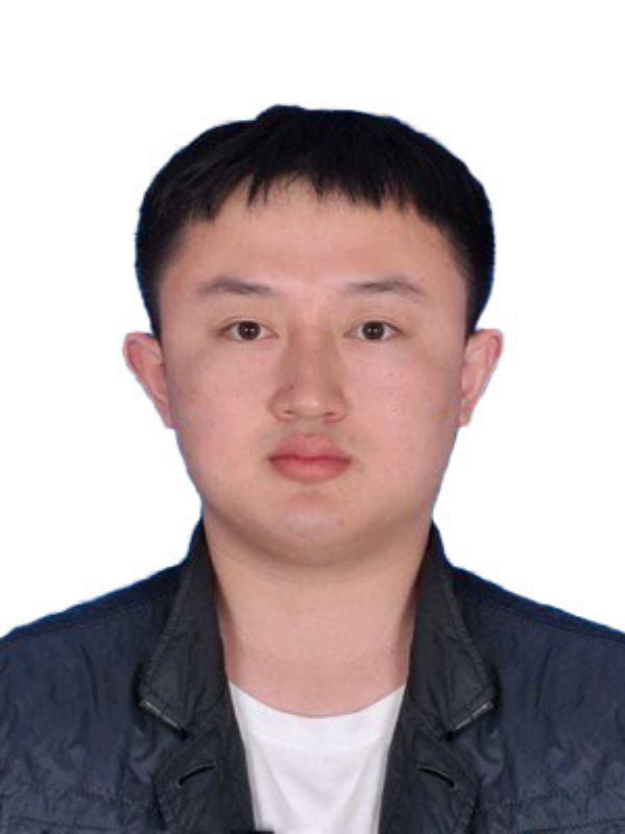}}]{Hao Wei}
is currently a Ph.D. candidate with the Institute of Artificial Intelligence and Robotics at Xi'an Jiaotong University, and a visiting Ph.D. student at the University of Western Australia, supervised by Professor Ajmal Mian. He received his B.Sc. and M.Sc. degrees from Yangzhou University and Nanjing University of Science and Technology in 2018 and 2021, respectively. His research interests include image deblurring, compression and rescaling, and other low-level vision problems.
\end{IEEEbiography}

\begin{IEEEbiography}
[{\includegraphics[width=1in,height=1.25in,clip,keepaspectratio]{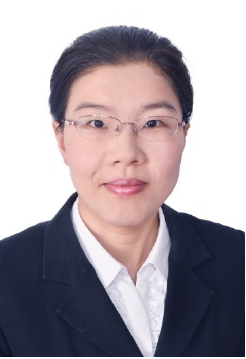}}]{Yanhui Zhou}
received the M.S. and Ph.D. degrees in electrical engineering from the Xi'an Jiaotong University, Xi'an, China, in 2005 and 2011, respectively. She is currently an associate professor with the School of Information and Telecommunication at Xi’an Jiaotong University. Her current research interests include image/video compression, computer vision and deep learning.
\end{IEEEbiography}

\begin{IEEEbiography}[{\includegraphics[width=1in,height=1.25in,clip,keepaspectratio]{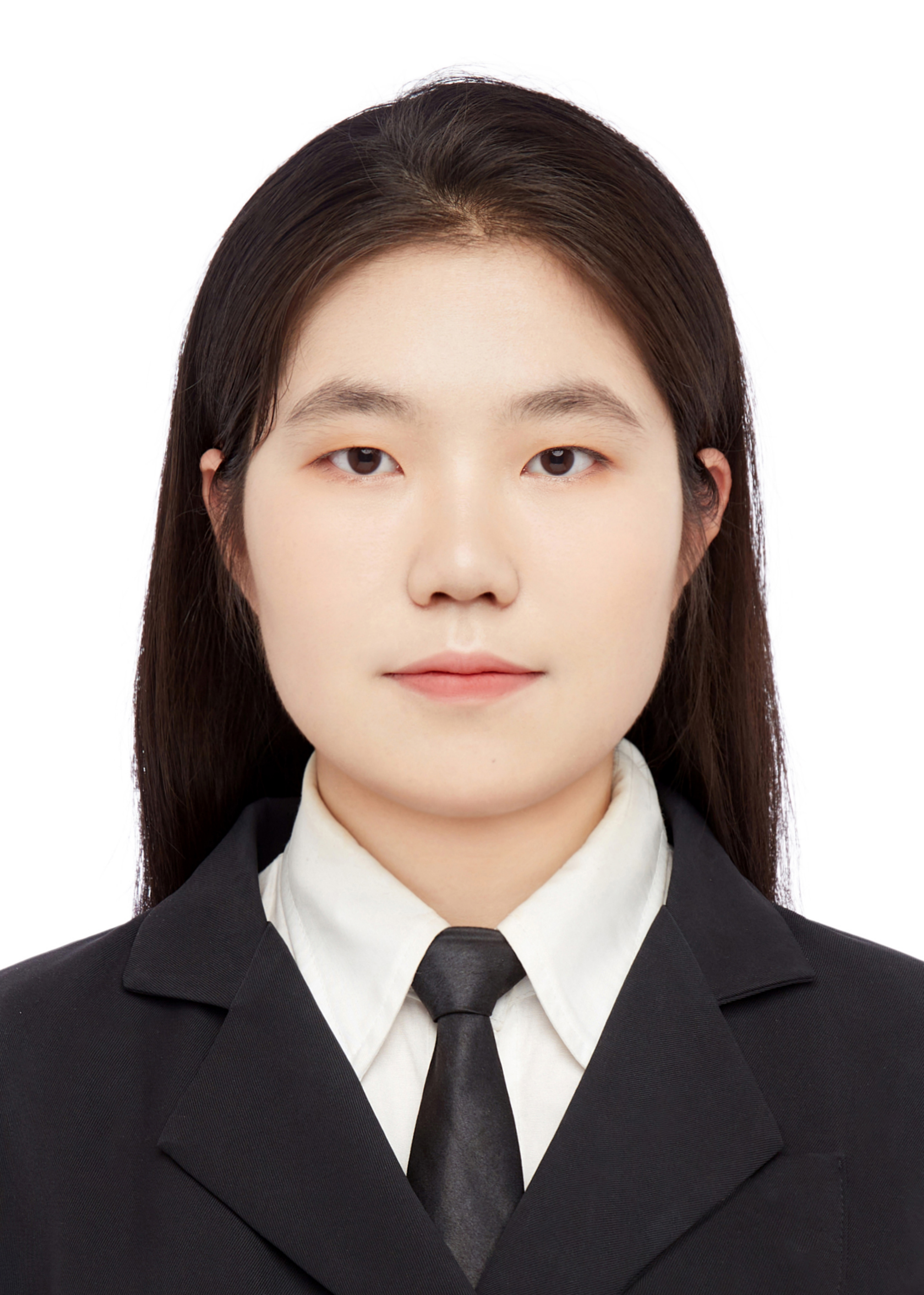}}]{Yiwen Jia}
is currently pursuing a master's degree at Xi'an Jiaotong University. She received her bachelor's degree from  Northwestern Polytechnical University in 2023. Her research interests include image compression and other visual problems.
\end{IEEEbiography}

\begin{IEEEbiography}[{\includegraphics[width=1in,height=1.25in,clip,keepaspectratio]{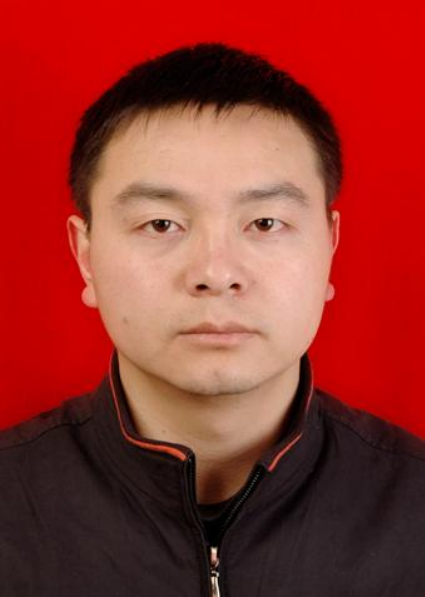}}]{Chenyang Ge}
is currently a professor at Xi'an Jiaotong University. He received the B.A., M.S., and Ph.D. degrees at Xi'an Jiaotong University in 1999, 2002, and 2009, respectively. His research interests include computer vision, 3D sensing, new display processing, and SoC design.
\end{IEEEbiography}

\begin{IEEEbiography}
[{\includegraphics[width=1in,height=1.25in,clip,keepaspectratio]{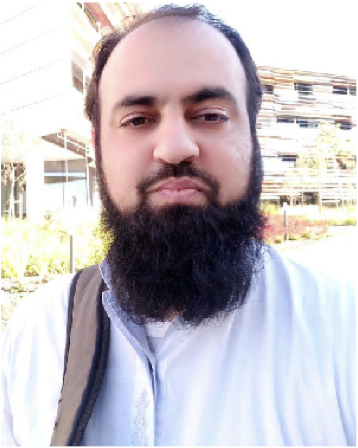}}]{Saeed Anwar} 
received the master’s degree (Hons.) from the Erasmus Mundus Vision and Robotics (Vibot), jointly offered by Heriot-Watt University, U.K., the University of Girona, Spain, and the University of Burgundy, France, and the Ph.D. degree from The Australian National University and National ICT Australia. He was with NICTA and CSIRO’s Data61, Australia. He is currently an Assistant Professor with the King Fahd University of Petroleum and Minerals (KFUPM),
Saudi Arabia. He also holds honorary positions with The Australian National University (ANU), The University of Technology Sydney (UTS), and the University of Canberra, Australia. He has a strong teaching experience in many reputed universities and a substantial industry presence. He leads commercial projects and supervises Ph.D., M.S., and B.S. students. He has published in top-tier conferences and journals, including One Best Paper Nomination in CVPR and a Best Paper Honorable Mention in Pattern Recognition (PR).
\end{IEEEbiography}

\begin{IEEEbiography}[{\includegraphics[width=1in,height=1.25in,clip,keepaspectratio]{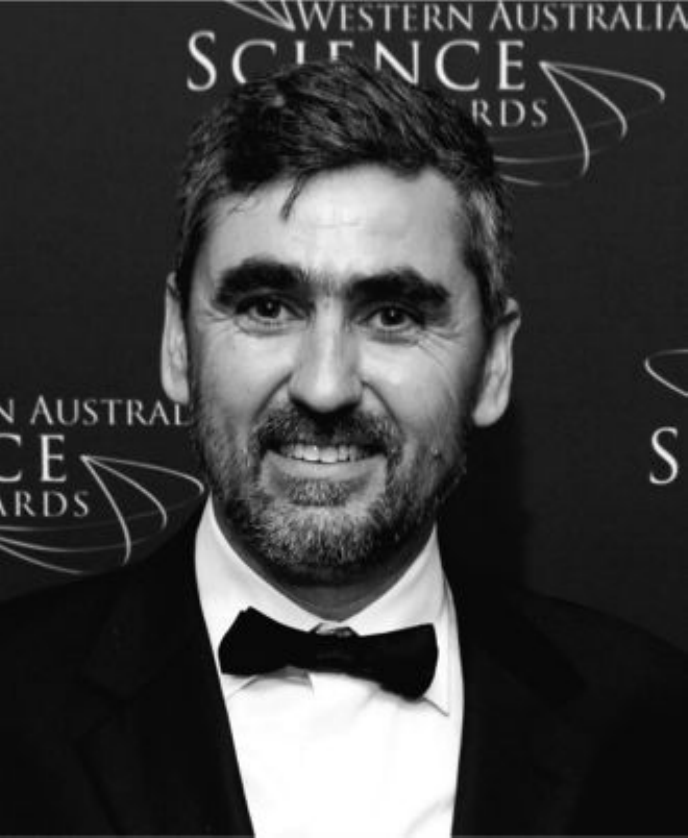}}]{Ajmal Mian}(Senior Member, IEEE)
is a Professor of Computer Science at The University of Western Australia. He is the recipient of three esteemed fellowships from the Australian Research Council (ARC). He has also received several research grants from the ARC, the National Health and Medical Research Council of Australia, US Department of Defense and the Australian Department of Defense with a combined funding of over \$41 million. He received the West Australian Early Career Scientist of the Year 2012 award and the HBF Mid-career Scientist of the Year 2022 award. He has also received several other awards including the Excellence in Research Supervision Award, EH Thompson Award, ASPIRE Professional Development Award, Vice-chancellors Mid- career Award, Outstanding Young Investigator Award, and the Australasian Distinguished Dissertation Award. He is an IAPR Fellow and Distinguished Speaker of the ACM. He also served as a Senior Editor for IEEE Transactions in Neural Networks and Learning Systems and Associate Editor for IEEE Transactions on Image Processing and the Pattern Recognition Journal. He was the General Co-Chair of DICTA 2019 and ACCV 2018. His research interests are in 3D computer vision, machine learning, and video analysis.
\end{IEEEbiography}